\documentclass{aa}
\usepackage{graphicx}
\usepackage{txfonts}
\newcommand{\msun}{~\mbox{M$_{\sun}$}}
\newcommand{\lsun}{~\mbox{L$_{\sun}$}}
\newcommand{\natd}[2]{\mbox{$#1 \cdot 10^{#2}$}}
\newcommand{\pder}[2]{\frac{\partial #1}{\partial #2}}
\newcommand{\pderr}[1]{\pder{#1}{R}}
\newcommand{\pdert}[1]{\pder{#1}{t}}
\newcommand{\pderphi}[1]{\pder{#1}{\varphi}}

\begin{document}
   \title{On stability and spiral patterns in polar disks}

   \subtitle{}

   \author{Ch.\ Theis
           \inst{1,2}
          \and
           L. Sparke\inst{3,4}
          \and
           J. Gallagher\inst{4}
          }

   \offprints{Ch.\ Theis}

   \institute{Institute of Theoretical Physics and Astrophysics,
              University of Kiel, 24098 Kiel, Germany
         \and
              Institute of Astronomy, University of Vienna,
              T\"urkenschanzstr.\ 17, A-1180 Vienna, Austria\\
              \email{theis@astro.univie.ac.at}
         \and    
                Max-Planck-Institute for Astrophysics,
                Postfach 1317, 
                85741 Garching, Germany 
         \and
             University of Wisconsin, Department of Astronomy, 475 N. Charter 
             St., Madison, WI 53706, USA\\
             \email{sparke@astro.wisc.edu, jsg@astro.wisc.edu}
             }

   \date{Received ; accepted }

   \abstract{To investigate the stability properties of polar disks we
     performed two-dimensional hydrodynamical simulations for flat
     polytropic gaseous self-gravitating disks which were perturbed 
     by a central S0-like component.
     Our disk was constructed to resemble that of
     the proto-typical galaxy NGC 4650A.  This central perturbation
     induces initially a stationary two-armed tightly-wound leading
     spiral in the polar disk.  For a hot disk (Toomre parameter
     $Q>1.7$), the structure does not change over the simulation time
     of 4.5 Gyr.  In case of colder disks, the self-gravity of the
     spiral becomes dominant, it decouples from the central
     perturbation and grows, until reaching a saturation stage in
     which an open trailing spiral is formed, rather similar to that
     observed in NGC~4650A.  The timescale for developing non-linear
     structures is 1-2 Gyr; saturation is reached within 2-3 Gyr. The
     main parameter controlling the structure formation is the Toomre
     parameter. The results are surprisingly insensitive to the
     properties of the central component.  If the polar disk is much
     less massive than that in NGC~4650A, it forms a weaker
     tightly-wound spiral, similar to that seen in dust absorption in
     the dust disk of NGC~2787. 
     Our results are derived for a polytropic equation of
     state, but appear to be generic as the adiabatic exponent is
     varied between $\gamma = 1$ (isothermal) and $\gamma = 2$ (very
     stiff).

      }
   
   \maketitle


\section{Introduction}

 Polar ring galaxies (PRGs) are characterized by an early-type
or elliptical central component and a gas-rich ring or disk which is
nearly polar with respect to the central galaxy. 
The spatial orientation and the
kinematics of the main components makes them ``multi-spin systems'' 
(Rubin \cite{rubin94}) with mutually orthogonal spin vectors. In the
polar-ring catalog (Whitmore et al.\ \cite{whitmore90}) about 100
PRGs or PRG candidates are given. The polar structures 
can be either narrow rings, or radially-extended disks. Both
morphologies are about equally frequent. 

A prototype of a PRG with a radially-extended disk is NGC
4650A. It is characterized by a central gas-poor S0 component and an
exponential polar disk much larger than the central galaxy. At
an adopted distance of 35 Mpc, the central component has a luminosity 
$L_I \approx \natd{3}{9} \lsun$, corresponding to a mass of
$\natd{5}{9} \msun$ in stars, while for the polar disk 
$L_I \approx \natd{1.2}{9} \lsun$, so the stellar
mass $\natd{3}{9} \msun$ (Gallagher et al.\ \cite{gallagher02}). 
The polar disk shows roughly an exponential distribution of stellar light 
(Iodice et al.\ \cite{iodice02}).  It is gas-rich, with an HI-mass
of about $\natd{7}{9} \msun$ at the adopted distance (Arnaboldi et al.\ 
\cite{arnaboldi97}), and about 10\% as much mass in molecular form
(Watson et al.\ \cite{watson94}).  
The HI maps show gas rotating at speed 120km\,s$^{-1}$ out to a radius
of 25~kpc, which corresponds to a dynamical mass of $\natd{8}{10} \msun$.
The polar disk is blue, with about 1\msun~yr$^{-1}$ 
of ongoing star formation, whereas in the S0 component star
formation appears to have
ceased $3-5 \,$Gyrs ago (Gallagher et al.\ \cite{gallagher02}).

Arnaboldi et al.\ pointed out the presence of two open
spiral arms in the polar disk, with approximate mirror symmetry 
about the central S0, in both optical and near-infrared light. 
These have a spatial extent of about 20 kpc in diameter 
(see their Fig.~10 or our Fig.\ \ref{n4650_m11_comp} below). 
The arms are also traced by a few blue 
knots, suggesting that they represent real density enhancements where
stars have formed, and not simply a warp in the polar disk (Gallagher
et al.\ \cite{gallagher02}).

The polar disk in NGC~2787, which is classified SB0/a with a LINER
nucleus, exhibits a very different morphology, with a tightly-wound
spiral in the inner 7~arcsec or 200~pc of the polar disk.  The HST/WFPC2
image (see Figure~13 of Erwin \& Sparke \cite{erwin03} or Figure~4 of Sil'chenko
\& Afansiev \cite{silchenko04}) shows the spiral in dust absorption SE of the
nucleus where the polar disk lies in front of the galaxy.
The central galaxy is about as luminous as NGC~4650A, with 
$L_B \approx \natd{2}{9} \lsun$ 
at a distance of 7.5~Mpc, and probably has a similar mass.
But the polar structure is much less massive, with only 
$\natd{3}{8} \msun$ of HI and H$_2$ (Welch \& Sage \cite{welch03}), 
and most of this is in a ring at radius $\sim 6.4$~arcmin or 14~kpc. 

Though polar ring galaxies (PRGs) are not very frequent (about 0.5\%
of all galaxies (Whitmore et al.\ \cite{whitmore90})), they attracted
the interest of astronomers mainly for two  
reasons. First, the fact that they are multi-spin systems
makes them prime targets to study rotation curves in orthogonal directions.
From that, the flattening of dark haloes might be deduced 
(e.g.\ Whitmore et al.\ \cite{whitmore87}, 
Sackett \& Sparke \cite{sackett90},
Sackett et al.\ \cite{sackett94}). 
Second, their formation mechanism is still under debate though
it is assumed that galaxy interactions play a key r\^ole. 
Basically two formation concepts have been discussed in the past. 
One scenario is a tidal accretion model, in which mass is
transferred in an encounter between two galaxies 
(e.g.\ Reshetnikov \& Sotnikova \cite{reshetnikov97}). The second
assumes a head-on merger between two orthogonal spiral galaxies 
(Bekki \cite{bekki98}). Bournaud \& Combes (\cite{bournaud03})
performed numerical simulations to compare the probability of both
scenarios, favouring the accretion 
scenario. However, a lot of assumptions which are observationally
hard to determine enter the estimate of the probabilities for 
both scenarios.  

In order to understand the formation and evolution of polar disks,
their ages and, thus, their stability properties are of special
interest.  From their regular morphology, many appear to be stable
on at least several dynamical timescales (see e.g.\ Sparke 
\cite{sparke2004} for a review).
If so, differential precession effects have to be suppressed 
effectively. Steiman-Cameron \& Durisen (\cite{steiman82}) suggested that a 
non-spherical halo might be a stabilizing factor. Sparke
(\cite{sparke86}) and Arnaboldi \& Sparke (\cite{arnaboldi94})
emphasized that self-gravity may stabilize polar disks. 
Especially, in massive polar disks like in NGC 4650A self-gravity
is expected to be important. 

  In this paper, we want to investigate the stability properties
of self-gravitating polar disks which are perturbed by a central
galactic component. Our numerical analysis is based on 
two-dimensional simulations for flat disks. Such calculations
have been widely used for modelling stellar and gaseous disks 
(e.g.\ Englmaier \& Shlosman \cite{englmaier00}, Orlova et al.\ 
\cite{orlova02}, Masset \& Bureau \cite{masset03}), 
for simulating star-forming disks 
(e.g.\ Korchagin \& Theis \cite{korchagin99}) or dusty nuclear disks 
(Theis \& Orlova \cite{theis04}). This numerical approach allows not only to 
study the evolution during the linear stage, but also to follow the 
system deeply into the non-linear regime. From that we get e.g.\ the
timescales of the different evolutionary stages or the spatial pattern
and kinematics of the system.

  The numerical code is described in Sect.\ \ref{hydrodynamiccode}.
In Sect.\ \ref{initialmodel} the initial model is presented.
The numerical simulations are given in Sect.\ \ref{sect_noperturbation}
and \ref{sect_perturbation}.
Finally, we discuss and summarize our results in Sect.\ \ref{discussion}.


\section{Hydrodynamic code}
\label{hydrodynamiccode}

\subsection{Numerical Method}

     In our simulations we solve the two-dimensional hydrodynamical equations
for a flat disk in cylindrical coordinates (with $z=0$). 
This represents a polar ring or disk around a galaxy that is
flattened in the $x-z$ plane, as shown in Figure \ref{geometry_scheme}.
In some images
cartesian coordinates are shown: then the positive $x$-axis corresponds to 
$\varphi=0$ and $\varphi$ increases counter-clockwise.
We solve the
\begin{enumerate}
  \item continuity equation for the surface density $\Sigma(R)$
     \begin{equation}
         \label{continuity}
         \pdert{\Sigma} + 
         \frac{1}{R} \pderr{ (R \Sigma u)} +
         \frac{1}{R} \pderphi{ (\Sigma v)} = 0
     \end{equation}
  \item the Euler equation for the radial velocity $u$
     \begin{equation}
         \label{euler_u}
         \pdert{u} +
         u \pderr{u} + 
         \frac{v}{R} \pderphi{u} -
         \frac{v^2}{R} +
         \frac{1}{\Sigma} \pderr{P} +
         \pderr{\Phi} = 0
     \end{equation}
  \item the Euler equation for the azimuthal velocity $v$
     \begin{equation}
        \label{euler_v}
        \pdert{v} + 
        u \pderr{v} +
        \frac{v}{R} \pderphi{v} +
        \frac{v u}{R} +
        \frac{1}{\Sigma} \pderphi{P} +
        \frac{1}{R} \pderphi{\Phi} = 0
     \end{equation}
\end{enumerate}
The gravitational potential $\Phi \equiv 
\Phi_{\rm disk} + \Phi_{\rm hb} + \Phi_{\rm bar}$
consists of three components: 
the polar disk, a halo/bulge, and a 
contribution from the flattened S0 galaxy.
The gravitational potential $\Phi_{\rm disk}$ of the flat 
polar disk is given by 
(Binney \& Tremaine \cite{binney87}, hereafter 
BT87, Sect.\ 2.8) 
\begin{eqnarray}
  \Phi_{\rm disk}(R,\varphi) & = & - G \int_0^\infty R^\prime dR^\prime \cdot
                     \nonumber \\
      & &        \int_0^{2\pi} 
               \frac{\Sigma(R^\prime,\varphi^\prime) d\varphi^\prime}
                    {\sqrt{{R^\prime}^2 + R^2 - 2 R R^\prime
                       \cos(\varphi^\prime - \varphi) }}  \, .
\end{eqnarray}
This sum is calculated using a FFT technique applying the 2D Fourier 
convolution theorem for polar coordinates. 
External potentials
describe the contribution of a rigid halo/bulge. The latter can be
done in two ways: either a halo/bulge mass distribution is given
from which the equilibrium rotation curve is derived (considering also
the pressure forces and the self-gravity of the disk) or the
rotation curve $v_c(R)$ is given and the halo/bulge mass distribution is
derived. In the latter case (which we used here), the contribution 
$\partial \Phi_{\rm hb} / \partial R$, i.e.\ the
enclosed mass of the halo/bulge component, is calculated from
\begin{equation}
   v_c^2(R) = R \pder{}{R}(\Phi + \Phi_{\rm hb}) \, .
   \label{eqvcphihalo}
\end{equation}
Though formally, any profile of $ \Phi_{\rm hb}$ might be a solution
of Eq.\ (\ref{eqvcphihalo}), we only accept physically meaningful, 
radially increasing cumulative mass distributions. 
More details of our setup of the
rotation curve can be found in Sect.\ \ref{sect_rotationcurve}.

In the $x-y$ plane of the polar material,
the oblate S0 galaxy is represented by the bar-like
potential of a stationary Ferrers' bar (BT87, Sect.\ 2.3). 
In order to avoid an initial ''kick'' on the
disk, the mass $M_{\rm bar}$ of the bar-like perturbation
is slowly enhanced starting from $M_{\rm bar} (t=0) = 0$.
During this adiabatic turn-on process the mass $M_{\rm hb,<a}$ 
of the halo/bulge inside the semi-major axis $a$ is reduced by keeping the
sum of the masses $M_{\rm bar}(t) + M_{\rm hb,<a}(t) = {\rm const}$.

For the pressure terms we applied a polytropic equation of state
\begin{equation}
   P = C \cdot \Sigma^\gamma
  \label{polytrope}  
\end{equation}
with an adiabatic exponent of $\gamma=5/3$. 
This allows to close the set of hydrodynamical equations with the 
equations for the first order moments, i.e.\ the Euler equations 
(\ref{euler_u}) and (\ref{euler_v}).
 The hydrodynamic equations (\ref{continuity}) -- (\ref{euler_v}) 
are discretized on a logarithmic Eulerian grid with a constant logarithmic
grid spacing. The different terms in the equations
(\ref{continuity}) -- (\ref{euler_v}) are taken into account by
applying an operator splitting technique like e.g.\ in the ZEUS
program (Stone \& Norman \cite{stone92}). Advection is performed by the
second-order van-Leer scheme. The timestep is determined according 
to the usual Courant-Friedrichs-Levy criterion. For our simulations we use 
reflecting boundary conditions in radial direction (and periodic in 
azimuthal direction).  
Our simulations are mainly done on a grid with 270x270 cells.

Most of the simulations are performed without any explicit artificial
viscosity (AV).  Test calculations with different amounts of AV
did not significantly differ from the reference simulation without
AV. During the phase of linear growth the global amplitudes show
almost no quantitative difference. Later, in the saturation stage, shocks 
become more important and AV gives different results. However, the 
saturation levels just scatter around the models without AV. Because of that
and because we are mainly interested in the growth time for an instability,
we did not apply AV in most of the simulations, though it might 
improve the treatment of shocks in the late saturation stage.


\subsection{Units}

  The units are chosen to be
  $10^9 \msun$ and 1 kpc for the mass and length scale, respectively.
  The gravitational constant was set to $G=1$. This results in a
  velocity unit of 65.6 km\,s$^{-1}$ and a unit for the 
  angular speed of 65.6 km\,s$^{-1}$\,kpc$^{-1}$. The time unit
  is 14.9 Myr. These units are used throughout the paper, unless
  other units are given explicitly.


\subsection{Dimensions, Times and Accuracy}

Motivated by the polar disk of NGC 4650A we adopted an inner
edge of 2 kpc for most of our simulations. The outer edge is set to
50 kpc to prevent perturbations generated at the 
outer edge from affecting the solutions in the ``central'' area of the
polar disk (10 kpc). When using a total of 270 radial grid cells,
our logarithmically equidistant grid has a radial cell size of 24 pc 
at the inner edge degrading to 0.6 kpc at the outer boundary.

A typical simulation run needs of the order of several $10^4$ to
$10^5$ timesteps, until the end of the calculation at \mbox{$t=300
  \approx 4.47$ Gyr} is reached. A typical timestep is of the order of
\mbox{$\Delta t \sim 1\dots 6 \cdot 10^{-3} \sim 1.5 \dots 9 \cdot
  10^4$ yr}.  The simulations were performed on a NEC-SX5 vector
computer.  The code is highly vectorized, reaching up to 50\% of the
peak performance of a single vector processor (sustained performance:
1.6-2 GFlop). The most time-consuming part of the simulations is the
calculation of the disk's self-gravity. Numerically this is done with
a fast Fourier transform from the MathKaisan library on the NEC-SX5
computer which is especially fast for this number of grid cells.  The
usual number in that resolution regime is $256=2^8$ cells per
dimension. However, the speed of the FFT becomes rather slow for this
array size due to bank conflicts in the memory access.

For the simulation of the unperturbed disk (model A) the energy was 
conserved to better than $5\cdot 10^{-8}$ (relative deviation), 
angular momentum better than $2\cdot 10^{-8}$ and the mass even within 
the machine accuracy of double-precision numbers. The accuracy of the
code decreases when the system becomes
non-linear and the saturation regime is reached. In that case the
accuracy is of percent level for both, energy and angular momentum.


\section{Initial model}
\label{initialmodel}

In order to mimic real polar disks to a reasonable degree the
parameters for our initial model were chosen to match closely the
observations of the prototype polar disk NGC4650A (Gallagher et al.\ 
\cite{gallagher02}).  However, we do not intend to have a
near-to-exact model for NGC4650A, but a ``prototypical'' polar disk.
Thus we preferred to describe the rotation curve or the mass
distribution by ``simple'' standard formulas instead of matching
exactly small- or intermediate-scale variations. We also do not
consider the multi-phase nature of the ISM or (energy feedback from)
star formation in our calculations.  In detail the model is
characterized by the mass distribution, the kinematics of the disk and
the applied equation of state.


\subsection{Polar disk}

In our standard model, a mass of $M_{\rm disk} = 1.2 \cdot 10^{10} \msun$
is distributed exponentially with a scale length of 4 kpc.  The
``punched'' disk has a sharp inner boundary at a radius of 2~kpc. 
The half-mass radius of the disk is about 7.3~kpc, and
90\% of its mass lies within the 'central' region at $R<16$~kpc.
Additionally, 
the surface density is slightly perturbed resulting in a random deviation 
from the mean surface density of an amplitude of $10^{-6}$. 

\subsection{Rotation curve}
\label{sect_rotationcurve}

\begin{figure}
   \resizebox{\hsize}{!}{
     \includegraphics[angle=270]{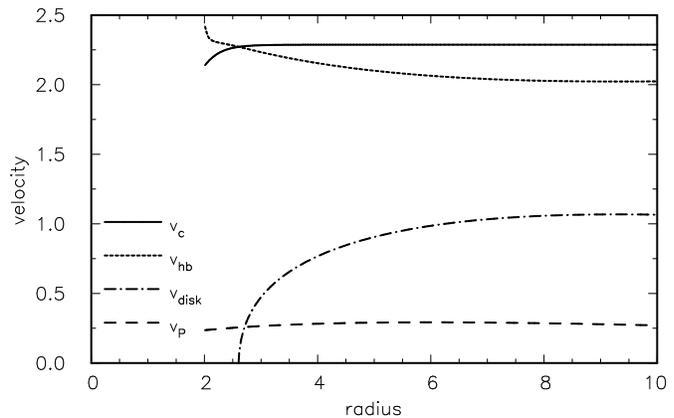}
   }
   \caption{Rotation curve $v_c$ and contributions of the different components
        ($v_\mathrm{hb}$: halo\&bulge, $v_\mathrm{disk}$:
   polar disk and $v_P$: pressure) for the unperturbed model A. Note
   that the contributions
         are added up quadratically for $v_c^2$.}
   \label{m12_vrot}
\end{figure}

  In principle our code allows us to adopt any prescribed rotation curve
by adjusting the halo/bulge mass distribution accordingly. In order to
determine the required halo/bulge potentials the gravity of the disk 
as well as the pressure gradient terms are taken into account.
They are subtracted from the rotation curve in order to get the 
remaining halo contribution. Though this
scheme allows always for a formal solution, only physical cases are accepted
by requiring that the enclosed halo/bulge mass is monotonically increasing
with radius. The halo/bulge potential is non-rotating.
The bar-like S0 component is treated as 
stationary except for a short, merely technical 
''adiabatic'' start phase (which is described below).

For the models in this paper we parametrized the rotation curve by
\begin{equation}
   v_c(R) = v_{\infty} \cdot 
              \frac{R/R_{\rm flat}}
                   {\left[ 1 + \left(R/R_{\rm flat} \right)^n \right]^{1/n}} \,\, .
\end{equation}
This guarantees a rigid rotation in the inner parts and a flat
rotation curve outside. The parameter $n$ defines the smoothness of the
transition between these two zones. For our simulations we mainly
used a sharp 
transition near $R_{\rm flat} = 2$ kpc by choosing $n=10$. Hence,
the rotation curve is basically flat all over the polar disk
(Fig.\ \ref{m12_vrot}). 

The initial azimuthal velocity of the gaseous phase
is calculated by the Jeans' equation for the radial velocity 
component which reads in cylindrical coordinates
\begin{equation}
   \pdert{u} + u \pder{u}{R} + \frac{v}{R} \pder{u}{\phi} 
              - \frac{v^2}{R} =
        - \frac{1}{\Sigma} \cdot \pder{P}{R}
        - \pder{}{R}\left( \Phi + \Phi_{\rm hb}\right) \,\, .
   \label{eqjeansu}
\end{equation}
\noindent
In case of the initial equilibrium, the radial velocity $u$ vanishes.
If we define the pressure contribution $v_P$ to the rotation
curve as 
\begin{equation}
   v_P^2 \equiv - \frac{R}{\Sigma} \cdot \pder{P}{R}
   \label{eqvpressure} \,\, ,
\end{equation}
then the azimuthal velocity $v$ is given by
\begin{equation}
   v^2 = 
           R \pder{}{R}\left( \Phi + \Phi_{\rm hb}\right)     
        + \frac{R}{\Sigma} \cdot \pder{P}{R} 
         = v_c^2(R) - v_P^2(R)   \,\, .
  \label{eqvazimuthal}
\end{equation}
\noindent
The rotation curve is
derived from the gravitational potential according to Eq.\
(\ref{eqvcphihalo}).  

Decomposing disk and halo contributions, we find that
our disk is submaximal by a factor of 3.6. This leaves sufficient 
mass for an additional central component.  It should be noted that
close to the inner edge the gravitational pull of the disk becomes negative, 
i.e.\ points outwards. This ''repulsive'' force has to be compensated for
by a halo contribution corresponding to a rotational velocity
exceeding the rotation curve.

We choose the rotation speed $v_{\infty}$ at large radii to be
150~km\,s$^{-1}$.  The rotation period at the inner edge is 5.9 time
units ($\approx$ 90 Myr) increasing to $\sim$ 300 Myr at the half-mass
radius of the disk and reaching $\sim$ 2 Gyr at its outer edge at 50
kpc. The maximum of \mbox{$\Omega(R)-\kappa(R)/2$} is important for
the existence of inner Lindblad resonances: here $\kappa$ is the
epicyclic frequency and $\Omega = v_c / R$ the circular speed.  This
maximum is reached at 2.55~kpc where \mbox{$\Omega-\kappa/2 = 0.236$}
or $=15.4$ km\,s$^{-1}$\,kpc$^{-1}$.  
In case of pure stellar disks, a two-armed spiral pattern
rotating at the angular rate $\Omega_p$ can propagate only where
\mbox{$|\Omega_p - \Omega| < \kappa/2$}, so any such spiral with
$\Omega_p < 0.236$ cannot propagate from the outer disk to the inner
boundary. In our study we deal, however, with gaseous disks, 
in which density waves can propagate as sound waves through the
Lindblad resonances, with little damping
(for a brief summary see 
Bertin (\cite{bertin00}), hereafter GB00, Chap.\ 16). 
Polar rings are extremely gas-rich; often more of their baryonic
mass is in cool gas than in stars (e.g.\ Sparke \cite{sparke2004}). 
So this is likely to be an appropriate approximation.


\subsection{Toomre parameter}

\begin{figure}
   \resizebox{\hsize}{!}{
     \includegraphics[angle=90]{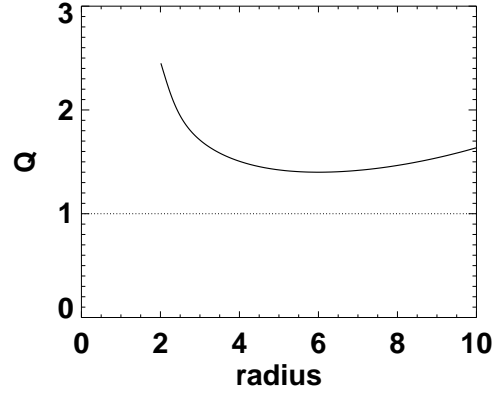}
   }
   \caption{Initial radial profile of the Toomre parameter $Q$ for the
      unperturbed model A and the reference model B. The minimum $Q$
      of 1.4 is reached at about 6 kpc.}
   \label{m12_toomre}
\end{figure}

  The local stability of a disk with respect to axisymmetric perturbations
is controlled by the Toomre parameter
\begin{equation}
  Q \equiv \frac{\sigma_R \kappa}{\pi G \Sigma} \,\, .
\end{equation}
$Q$ compares the stabilizing factors -- velocity dispersion $\sigma_R$
(sound speed) and radial oscillation frequency (or epicyclic frequency
$\kappa$) -- with the destabilizing
r\^ole of the self-gravity of a mass element with surface density
$\Sigma$ (e.g.\ BT87 Sect.\ 6.2). 
A value of $Q$ less than unity means instability 
to axisymmetric perturbations.  In gaseous disks the most 
unstable wavelength $\lambda_m$ in the radial direction is given by
\begin{equation}
  \lambda_m \equiv \lambda_{\rm crit} / 2
\end{equation}
with the critical wavelength $\lambda_{\rm crit}$ defined as 
\begin{equation}
  \lambda_{\rm crit} \equiv \frac{4 \pi^2 G \Sigma}{\kappa^2} \,\, .
\end{equation}

From numerical experiments it is known that disks become rather stable 
even to non-axisymmetric modes for minimum $Q$-values above 1.4. 
In case of flat rotation curves one can derive from linear stability
analysis that $Q>\sqrt{3} \approx 1.73$ guarantees 
that a fluid disk is stable 
with respect to all perturbation modes
(see e.g.\ Bertin et al.\ (\cite{bertin89}), 
Polyachenko et al.\ (\cite{polyachenko97})
or Section 15.2 of GB00).
This is also in good agreement with multi-component disks studied 
by Orlova et al.\ (\cite{orlova02}).
In our reference model the constant $C$ of Eq.\ (\ref{polytrope}) is 
adjusted in such a way that the initial minimum value of $Q$ is 1.4. 
This value is reached at \mbox{$R \approx 6$ kpc}.
Fig.~\ref{m12_toomre} shows the radial run of $Q$ for this model.


\begin{table}
  \caption{Properties of the numerical models}
  \label{table_models}
  \begin{tabular}{c|c|c|c}
    model & $M_{\rm bar} (\msun) $ & $Q_{\rm min}$ & comment \\ \hline
     A    &    --        & 1.4 & no central component \\ \hline
     B    & \natd{5}{9}  & 1.4 & reference model \\
          &              &     &  (with central component) \\ \hline
     MB1  & \natd{1}{9}  & 1.4 & small S0 component \\
     MB2  & \natd{1}{10} & 1.4 & ''maximum'' S0 component \\
     MB3  & \natd{5}{9}  & 1.4 & $n_b=-1$  non-homogeneous S0 comp.\\
     MB4  & \natd{5}{9}  & 1.4 & $n_b=-2$, ''isothermal'' S0 comp.\\ \hline
     T1   & \natd{5}{9}  & 1.4 & $t_{\rm adi} = 100$, long switch-on \\
     T2   & \natd{5}{9}  & 1.4 & temporary S0 comp. (off at $t=100$) 
              \\ \hline
     Q1   & \natd{5}{9}  & 1.3 & varied Toomre parameter \\
     Q2   & \natd{5}{9}  & 1.5 &  \\
     Q3   & \natd{5}{9}  & 1.6 &  \\
     Q4   & \natd{5}{9}  & 1.7 &  \\
     Q5   & \natd{5}{9}  & 1.8 &  \\
     Q6   & \natd{5}{9}  & 1.9 &  \\ \hline
          &              &     & varied mass of the polar disk \\
     MD1  & \natd{5}{9}  & 1.4 & \natd{6.0}{9} $\msun$ \\
     MD2  & \natd{5}{9}  & 1.4 & \natd{6.0}{9} $\msun$, but $Q_{\rm min}=1.2$ \\
     MD3  & \natd{5}{9}  & 1.4 & \natd{2.4}{10} $\msun$ \\
     MD4  & \natd{5}{9}  & 1.4 & \natd{3.7}{10} $\msun$, maximum disk \\ \hline
     R1   & \natd{5}{9}  & 1.4 & varied rotation profile, n=6\\
     R2   & \natd{5}{9}  & 1.4 & n=3\\ \hline
     S1   & \natd{5}{9}  & 1.4 & varied Eq.\ of state, $\gamma=2.0$\\
     S2   & \natd{5}{9}  & 1.4 & $\gamma=1.0$ (isothermal)\\
     S3   & \natd{5}{9}  & 2.1 & $\gamma=1.0$ (isothermal), $Q_{\rm min}=2.1$ \\ \hline
     G1   & \natd{5}{9}  & 1.4 & larger grid (686x686) \\
     G2   & \natd{5}{9}  & 1.4 & inner edge at $R_\mathrm{min} = 1$ kpc \\
     G3   & \natd{5}{9}  & 1.4 & as G2, but smooth transition of $\Sigma$\\
     G4   & \natd{5}{9}  & 1.2 & as G3, but reduced $Q_\mathrm{min}$\\ \hline
  \end{tabular}
\end{table}

\section{Models without an S0-like central component}
\label{sect_noperturbation}

  The aim of this paper is to study the stability of polar disks
perturbed by a central S0-like component. As a test of the code, but also 
for comparison with the models of the following sections, we first
describe in Sect.\ \ref{sect_noperturbation} the evolution of an 
unperturbed gaseous disk (model A). In Sect.\ \ref{sect_perturbation} we
perturb the polar disk by a central S0-like component. A list of the
models described in this paper is given in Table \ref{table_models}.

\begin{figure}
   \resizebox{\hsize}{!}{
     \includegraphics[angle=90]{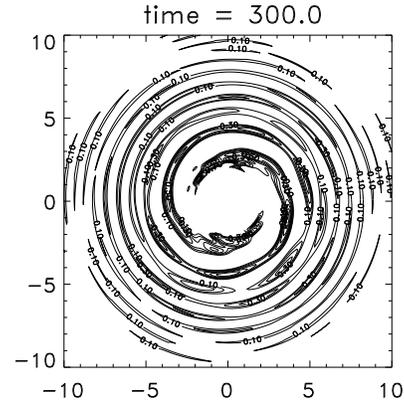}
   }
   \caption{Perturbation of the surface density profile at the end of
the simulation of model A ($t=300$). 
Contour lines are given in $\msun\,{\rm pc}^{-2}$ 
starting from 0.1 to 0.9 $\msun\,{\rm pc}^{-2}$ with a difference of 0.2 
$\msun\,{\rm pc}^{-2}$ between the contour lines.
For comparison, the unperturbed surface density at a distance of 5 kpc is 
37.2 $\msun\,{\rm pc}^{-2}$.}
   \label{m12_denspert_t300}
\end{figure}

   In the first set of simulations, we investigate
the evolution of an isolated galactic disk which has a central hole as described 
in the previous section. Such a disk is characterized by an exponential mass
profile, a basically flat rotation curve and a Toomre parameter of
$Q=1.4$, a value slightly above the one required for stability against 
axisymmetric perturbations.

   Inspecting the Lagrange radii, i.e.\ radii containing a constant
mass fraction, we see that the system is radially stable: 
the relative changes of the Lagrange radii are less
than 0.01\%  over the whole integration time of \mbox{$t=300=4.47$ Gyr}. 
On the other hand, 
the surface density profile shows the development of a pronounced two-armed 
trailing spiral structure until the end of the simulation 
(Fig.~\ref{m12_denspert_t300}). The spirals 
are tightly wound in agreement with a short critical wavelength 
$\lambda_{\rm crit}$ of about 3 to 4 kpc.
The most unstable wavelength is then about 
\mbox{$\lambda_m \sim 1.5 - 2 \,\, {\rm kpc}$} for a gaseous disk
(BT87) which reproduces roughly the wavelengths visible in 
Fig.~\ref{m12_denspert_t300}. The maximum amplitudes of the perturbed density
\mbox{$\Delta \Sigma (R,\varphi,t) \equiv \Sigma(R,\varphi,t) - 
\Sigma(R,\varphi,t=0)$} are below 1 $\msun\,{\rm pc}^{-2}$. Normalized to
the initial densities, $\Delta \Sigma / \Sigma$ is much less than 1\%, 
which would not be detectable in observations of real galaxies.

   Another way to visualize the nature and growth of instabilities in
disks is to consider the azimuthal Fourier modes $a_m$ at different radii.
They are calculated for $m \neq 0$ by
(cf.\ also Laughlin et al.\ \cite{laughlin98})
\begin{equation}
   a_m(R,t) \equiv \frac{1}{2\pi} 
           \int_0^{2\pi} \Sigma(R,\varphi,t) e^{im\varphi} d\varphi  \,\, .
\end{equation}
Their phase angle is then given by
\begin{equation}
    \phi_m(R,t) \equiv \tan^{-1} \left\{ 
                       \frac{\Im[a_m(R,t)]}{\Re[a_m(R,t)]}
                             \right\}
\end{equation}
This phase angle is a measure for the spatial orientation 
of the related perturbation. E.g.\ a bar described by
$\Sigma(R,\varphi) = \Sigma_0(R) \cdot \{1 + A \cos [2 (\varphi-\alpha)]\}$ 
has its major axis along $\varphi=\alpha$ or $\varphi=\alpha+\pi$, 
respectively. The corresponding phase angle for the $m$-mode is
$\phi_{m}(R) = m \alpha$.
Thus, the pattern speed for a mode $m$ can be calculated by
\begin{equation}
   \Omega_p(R,t) \equiv \frac{1}{m} \frac{d\phi_m(R,t)}{dt} \,\, .
   \label{eq_localpatternspeed}
\end{equation}
A measure for the global structure of a disk
are the (radially integrated) global Fourier modes defined by
\begin{equation}
   C_m(t) \equiv \frac{2\pi}{M_{\rm disk}} 
     \int_{R_{\rm low}}^{R_{\rm hig}} a_m(R,t) R dR \,\, .
\end{equation}
$M_{\rm disk}$ is the mass of the disk within the radial range 
$[R_{\rm low},R_{\rm hig}]$. The global Fourier amplitude is
the modulus of $C_m$, whereas the global growth rate is given
by the time derivative of the logarithmic Fourier amplitude, i.e.\ 
\begin{equation}
  \gamma_m \equiv \frac{d (\ln |C_m(t)|)}{dt} \,\, .
\end{equation}
It is also common to define a $m=0$-mode which denotes the temporal
evolution of the radial mass redistribution:
\begin{equation}
   C_0 \equiv \frac{2\pi}{M_{\rm disk}} \int_{R_{\rm low}}^{R_{\rm hig}}
         |\bar{\Sigma}(R,t) - \bar{\Sigma}_0(R)| R dR \,\, .
\end{equation}
We adopted $R_{\rm low} = 2$~kpc and $R_{\rm hig} = 50$~kpc,
integrating over the entire disk.
$\bar{\Sigma}(R,t)$ and $\bar{\Sigma}_0(R)$ are the azimuthally averaged
surface densities at time $t$, or at the start of the simulations, respectively.
As long as the evolution is well described by a linear approach, 
no radial mass transport should occur. Thus, the increase of $C_0$ is
a good indicator for the onset of non-linear effects.

\begin{figure}
   \resizebox{\hsize}{!}{
     \includegraphics[angle=270]{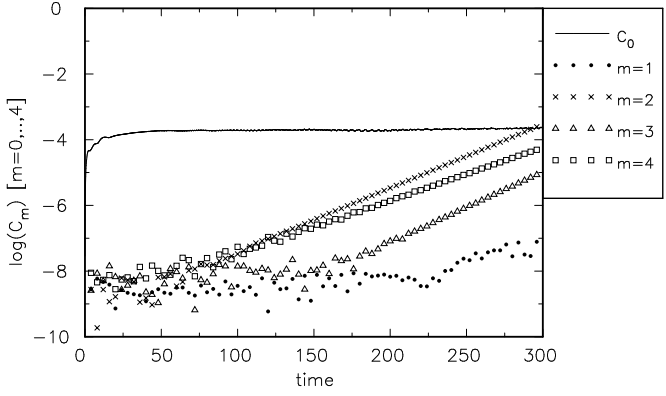}
   }
   \caption{Temporal evolution of the global Fourier modes for $m=1,\dots,4$
     of model A. 
     The mode $m=0$ describes the radial matter redistribution. For details
     see text.}
   \label{m12_modes}
\end{figure}

   Fig.~\ref{m12_modes} shows the temporal evolution of the global 
Fourier modes for the ''punched'' disk without a central component.
Starting from a low noise level, the even modes $m=2,4$ are the dominant
ones describing a basically two-armed structure. In agreement with predictions
of linear stability analysis they grow 
exponentially on a similar, but long timescale of
\mbox{$\gamma_m^{-1} \sim 20 \approx 300$ Myr}. In a late stage the
$m=3$-mode grows, too, whereas the $m=1$-mode remains stable throughout the
whole simulation. Until the end of the simulation the system never
reaches the regime of non-linear growth or saturation. The amplitudes
reached after 4.4 Gyr are so low that in real galaxies they would be 
practically invisible.

The $C_0$-amplitude shows a small initial
radial mass redistribution which is caused by the small deviations from
equilibrium due to the imprinted initial density perturbation. After quickly 
establishing a new equilibrium, the system remains radially stable.

\begin{figure}
   \resizebox{\hsize}{!}{
     \includegraphics[angle=90]{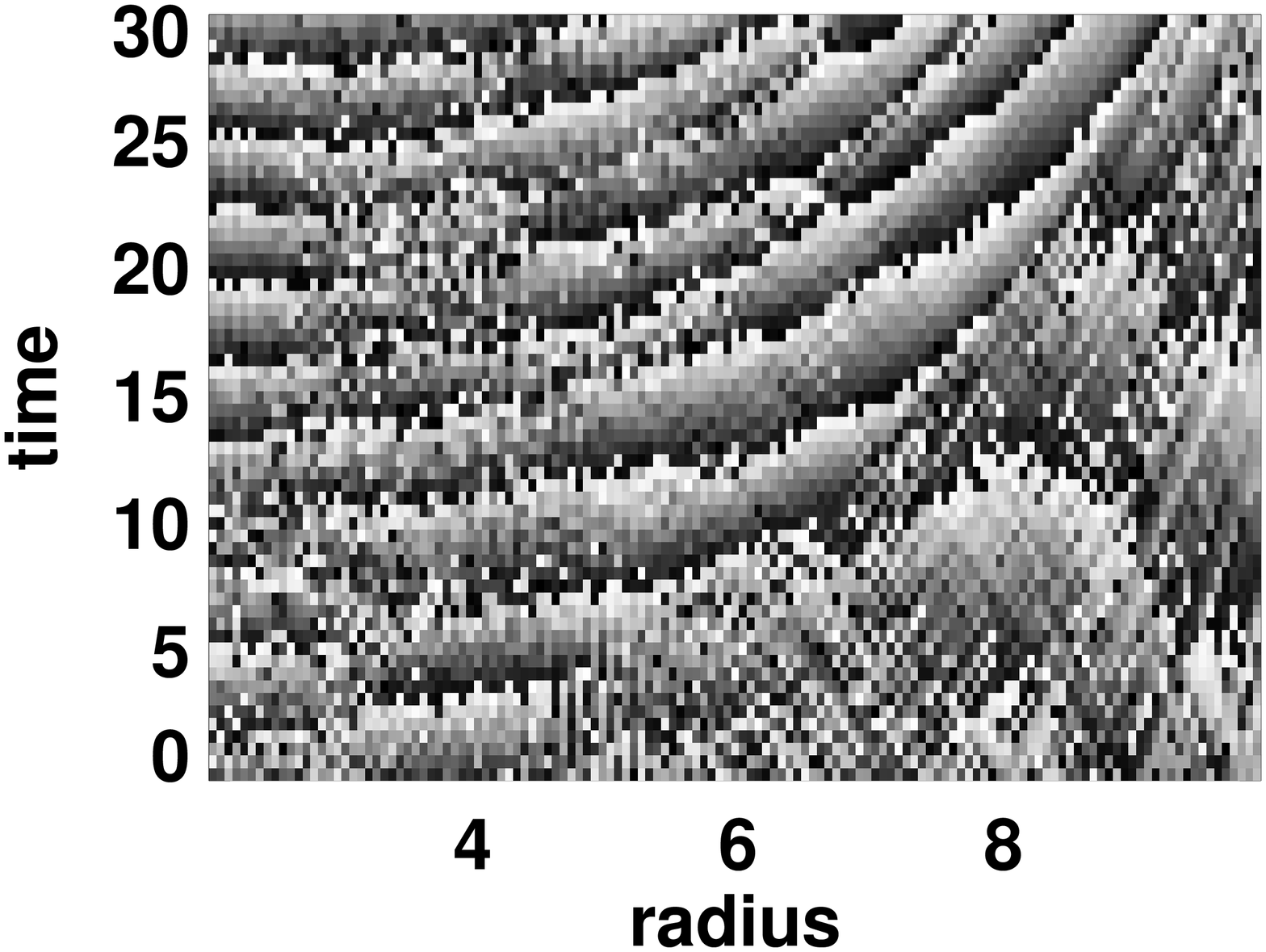}
   }
   \resizebox{\hsize}{!}{
     \includegraphics[angle=90]{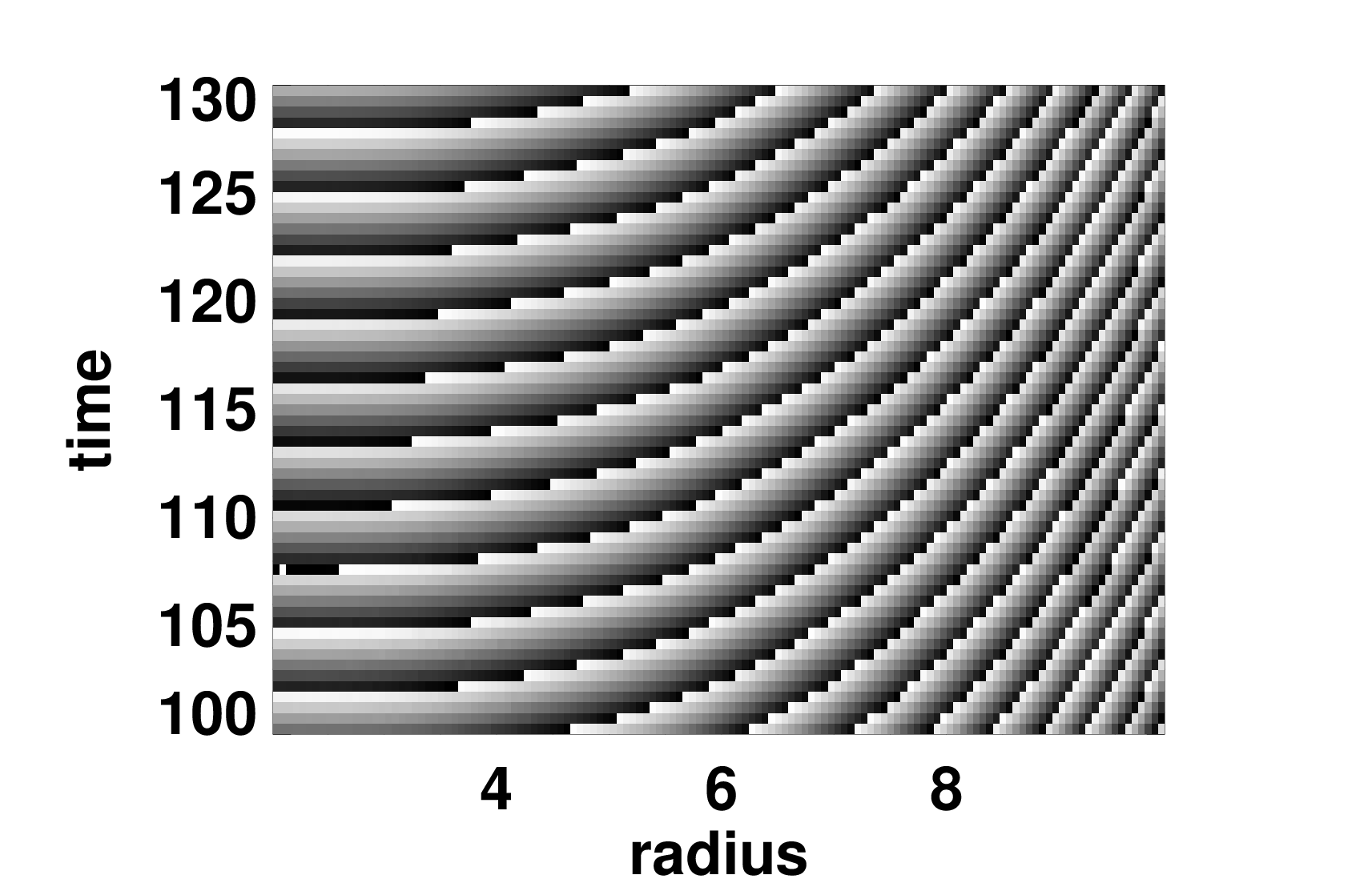}
   }
   \caption{Evolution of the phase angle $\phi_m$ of model A 
     for the $m=2$ mode during the 
     initial stage (upper diagram) 
     and the growing stage (lower diagram) 
     (0$^\circ$: white; 360$^\circ$: black).}
   \label{m12_phaseangle}
\end{figure}

  The evolution of the phase angles gives more detailed information on
the structure formation. During the initial stage of the simulation
no large scale pattern is discernible.
However, after time $t \sim 10$ coherent features are detectable for the
phase angle of the $m=2$ mode which become more and more
prominent indicating the weak two-armed structure of the system
(Fig.~\ref{m12_phaseangle}, upper diagram). The structure grows from inside
out, starting at a distance of about 4 kpc, well outside the
inner edge of the computational grid. Therefore, it is very unlikely
that the growing instability is caused by the sharp cut-off due to the
central hole of the disk or by boundary effects.

\begin{figure}
   \resizebox{\hsize}{!}{
     \includegraphics[angle=90]{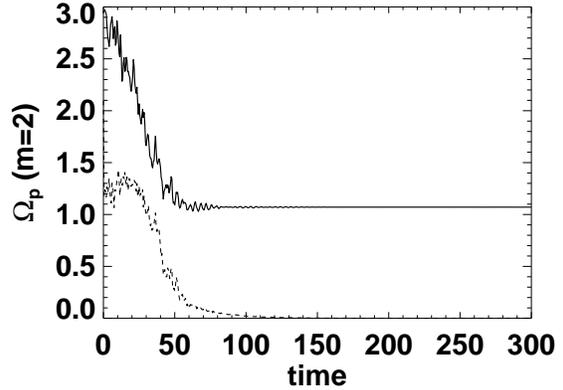}
   }
   \caption{Evolution of the radially averaged ($R< 10 \mathrm{kpc}$) pattern speed 
     of the $m=2$ mode (solid) and its mean deviation (dashed, lower
     curve) for model A.}
   \label{m12_m2patternspeed}
\end{figure}

In a later stage when the linear growth is well established the
tightly wound spiral structure is clearly visible in the phase angles
(Fig.~\ref{m12_phaseangle}, lower diagram). These allow an accurate
determination of the intrinsic pattern speed of the two-armed spiral
in the late stage as \mbox{$\Omega_{p,m=2} \approx 1.07 \approx 70$
  km\,s$^{-1}$kpc$^{-1}$} (Fig.~\ref{m12_m2patternspeed}).
$\Omega_{p,m=2}$ is the radial average of the local pattern speed
$\Omega_p(R,t)$ within the innermost 10 kpc.  The small mean
deviations of $\Omega_p(R,t)$ within this region indicate also the
formation of a large-scale $m=2$ pattern after $t=50$. For this
pattern speed, there are no inner Lindblad resonances and corotation
is close to but inside the inner boundary of the grid. Similar to the
$m$=2 mode, regular patterns are also found for the $m$=3- and
$m$=4-modes. Their pattern speeds are slightly smaller than that of
the $m$=2 mode (by 4\% and 7\%, respectively).
However, they are still too large to yield corresponding ILRs.


\section{Models with a central perturbation}
\label{sect_perturbation}

\begin{figure}
   \centerline{\resizebox{4cm}{!}{
     \includegraphics[angle=0]{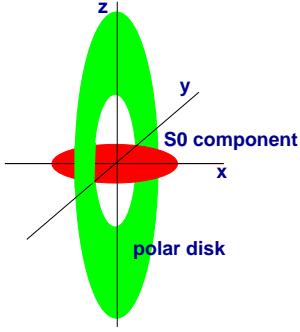}
      }}
   \caption{Schematic view of the geometry of the simulated polar disk.}
   \label{geometry_scheme}
\end{figure}

    In this section we describe the evolution of ''punched'' disks 
perturbed by a central S0- or elliptical-like component as a model 
for polar disks. 

The perturbation 
due to the S0 galaxy is given by the stationary potential 
of a Ferrers' bar which has a density distribution (BT87, Sect.\ 2.3)
\begin{equation}
     \rho_b(m^2) = \rho_0 \cdot \left( 1 - m^2 \right)^{\displaystyle n_b}  
         \cdot \Theta(1-m) \,\, .
\end{equation}
$\Theta$ is here the Heaviside function and
$m^2 = (x/a)^2 + (y/b)^2 + (z/c)^2 $ a modified radial coordinate.
For most of our simulations we choose $n_b=0$ assuming a homogeneous
mass distribution.  In our reference model,
the total mass of the central component is set to 
$M_{\rm bar} = 5 \cdot 10^9 \,\, \msun$. 

The S0 galaxy's major axis has a radial extent of
1 kpc, so the S0 disk lies completely inside the central hole of the 
polar disk. The axis ratios are set to $b/a = 1$ and $c/a = 0.4$ 
corresponding to an oblate spheroid. The minor axis points to the $z$-axis 
of the coordinate system. Thus, the ''disk'' of the central component is 
located in the $x$-$y$-plane and the polar disk is in the
$y$-$z$-plane. Hence, the disk plane of the central component
corresponds to the abscissa of the polar disk snapshots, while the
axis of the ordinate 
points to the $z$-axis, perpendicular to the disk plane of the
central S0 component (as shown in Fig.\ \ref{geometry_scheme}).

  In order to avoid switch-on effects, the 
bar-like potential of the central S0 component is turned on
slowly on a timescale of \mbox{$t_{\rm adi} = 20 \approx 300$ Myr}. 
During that time the mass of the S0 component 
is increased, while the same 
amount of mass is subtracted from the bulge/halo contribution. 


\subsection{The reference model}

\begin{figure}
   \resizebox{\hsize}{!}{
     \includegraphics[angle=270]{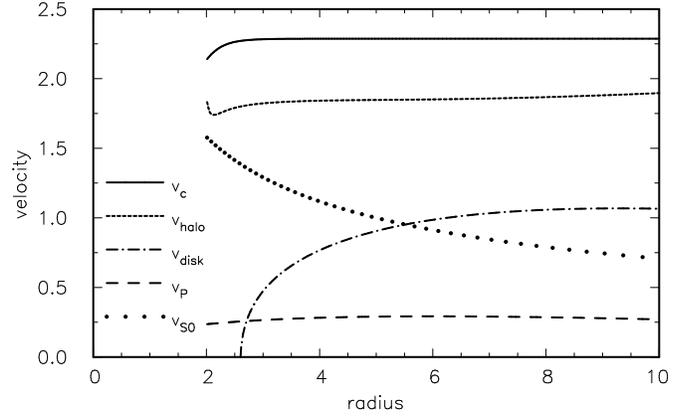}
   }
   \caption{Rotation curve $v_c$ of a polar disk including a central component 
         (model B). Shown are the contributions of the different components
         ($v_\mathrm{halo}$: halo, $v_\mathrm{disk}$: disk, $v_P$: pressure
         and $v_\mathrm{S0}$: S0 component).}
   \label{m11_vrot}
\end{figure}

  In order to start with a model resembling closely model A 
without a central component, we kept the rotation curve and the mass
distribution in the polar disk constant, but added the central 
perturbation with $M_{\rm bar} = 5 \cdot 10^9 \,\, \msun$ 
(model B). Due to the additional gravitational force 
generated by the central component, less mass is now in the halo. 
However, the velocity curve is still halo-dominated (Fig.\ \ref{m11_vrot}).

\begin{figure}
   \resizebox{\hsize}{!}{
     \includegraphics[angle=270]{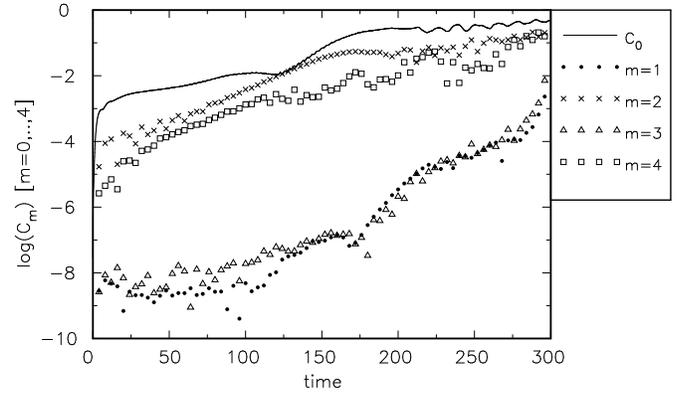}
   }
   \caption{Temporal evolution of the global modes of model B 
     for $m=1,\dots,4$. 
     The mode $m=0$ describes the radial matter redistribution.}
   \label{m11_modes}
\end{figure}

\begin{figure}
   \resizebox{\hsize}{!}{
     \includegraphics[angle=270]{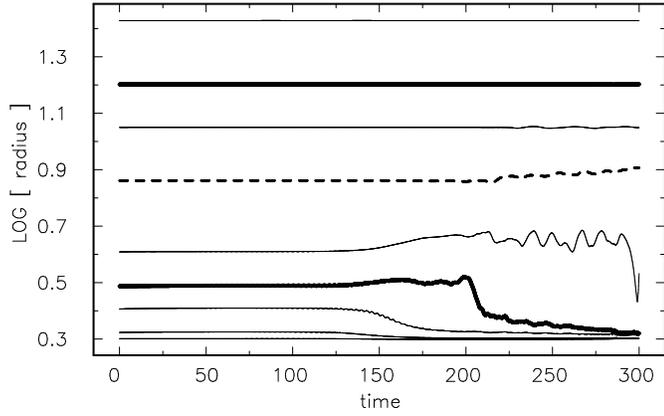}
   }
   \caption{Temporal evolution of the Lagrange-radii containing 0.1\%, 1\%,
       5\%, 10\% (thick line), 20\%, 50\% (dashed), 75\%, 90\% (thick)
       and 99\% of the disk mass for model B.}
   \label{m11_lagrange}
\end{figure}

\begin{figure}
   \resizebox{\hsize}{!}{
     \includegraphics[angle=90]{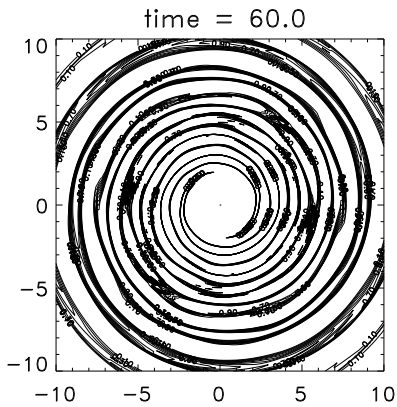}
   }
   \caption{Perturbation of the surface density profile of model B during the 
linear regime ($t=60=890$ Myr). Contour lines are given in 
$\msun$\,pc$^{-2}$ starting from 0.1 to 0.9 $\msun$\,$\mathrm{pc}^{-2}$ with a 
difference of 0.2 $\msun$\,$\mathrm{pc}^{-2}$ between the contour lines.
The surface density at 10 kpc is about $10 \msun$\,$\mathrm{pc}^{-2}$. Note the
leading arms of the formed pattern.}
   \label{m11_denspert_t60}
\end{figure}

\begin{figure}
   \resizebox{\hsize}{!}{
     \includegraphics[angle=90]{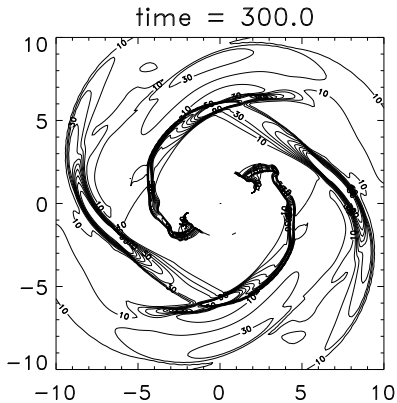}
   }
   \caption{Surface density profile in the saturation stage
     at the end of the simulation of model B ($t=300=4.47$ Gyr).
     Contour lines are given in $\msun$\,pc$^{-2}$ starting from 10 to 100 
     $\msun$\,$\mathrm{pc}^{-2}$ with a difference of 
     10 $\msun$\,$\mathrm{pc}^{-2}$ between the contour lines.
     The maximum relative density perturbation $\Delta \Sigma / \Sigma_0$ 
     is about 5.5 in the knot at (0,-6)and 11.6 near the center.}
   \label{m11_dens_t300}
\end{figure}

\begin{figure}
   \begin{center}
     \begin{minipage}{15cm}
        \includegraphics[angle=110,width=5cm]{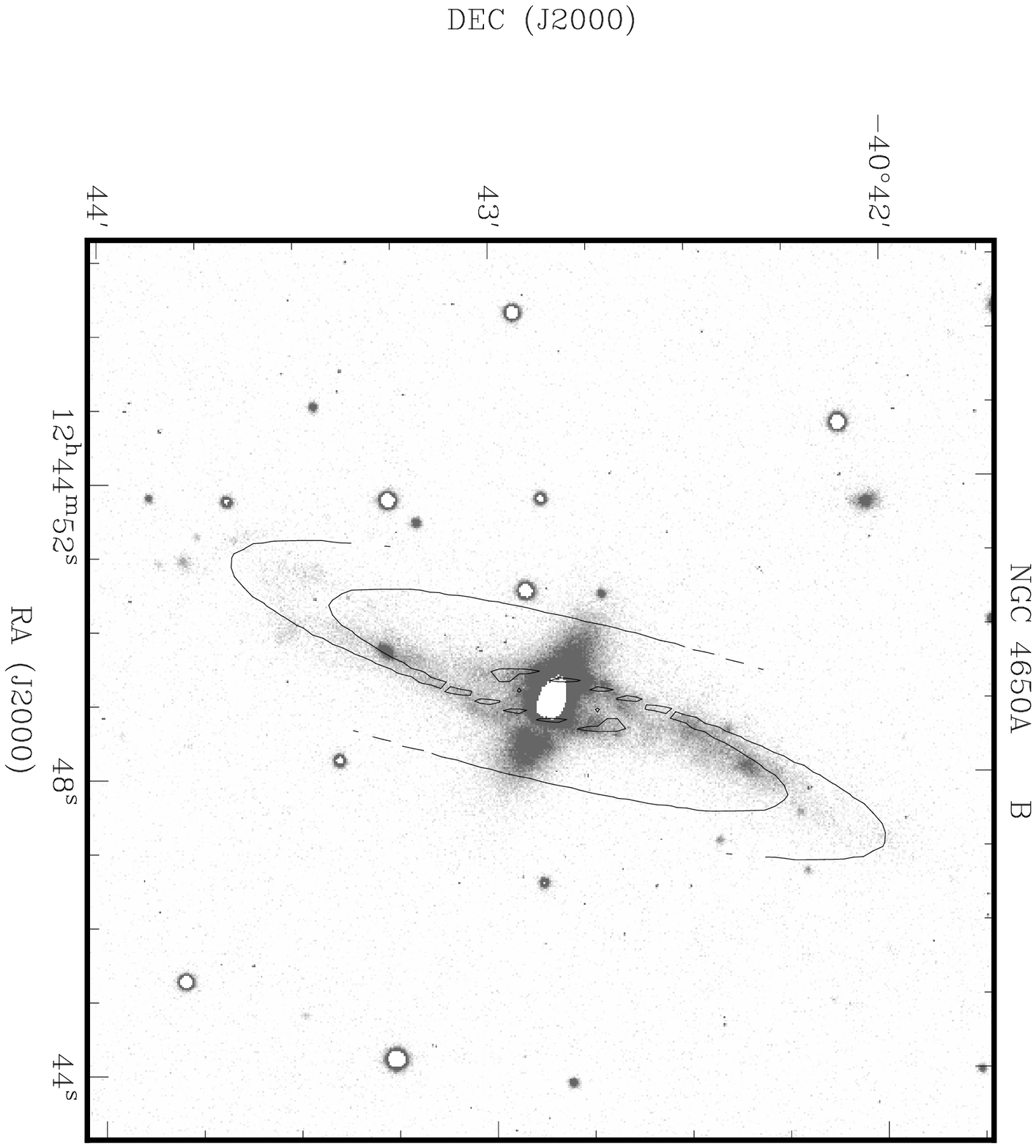}
        \hspace*{-2cm}\vspace*{0.5cm}
        \includegraphics[angle=90,width=7cm]{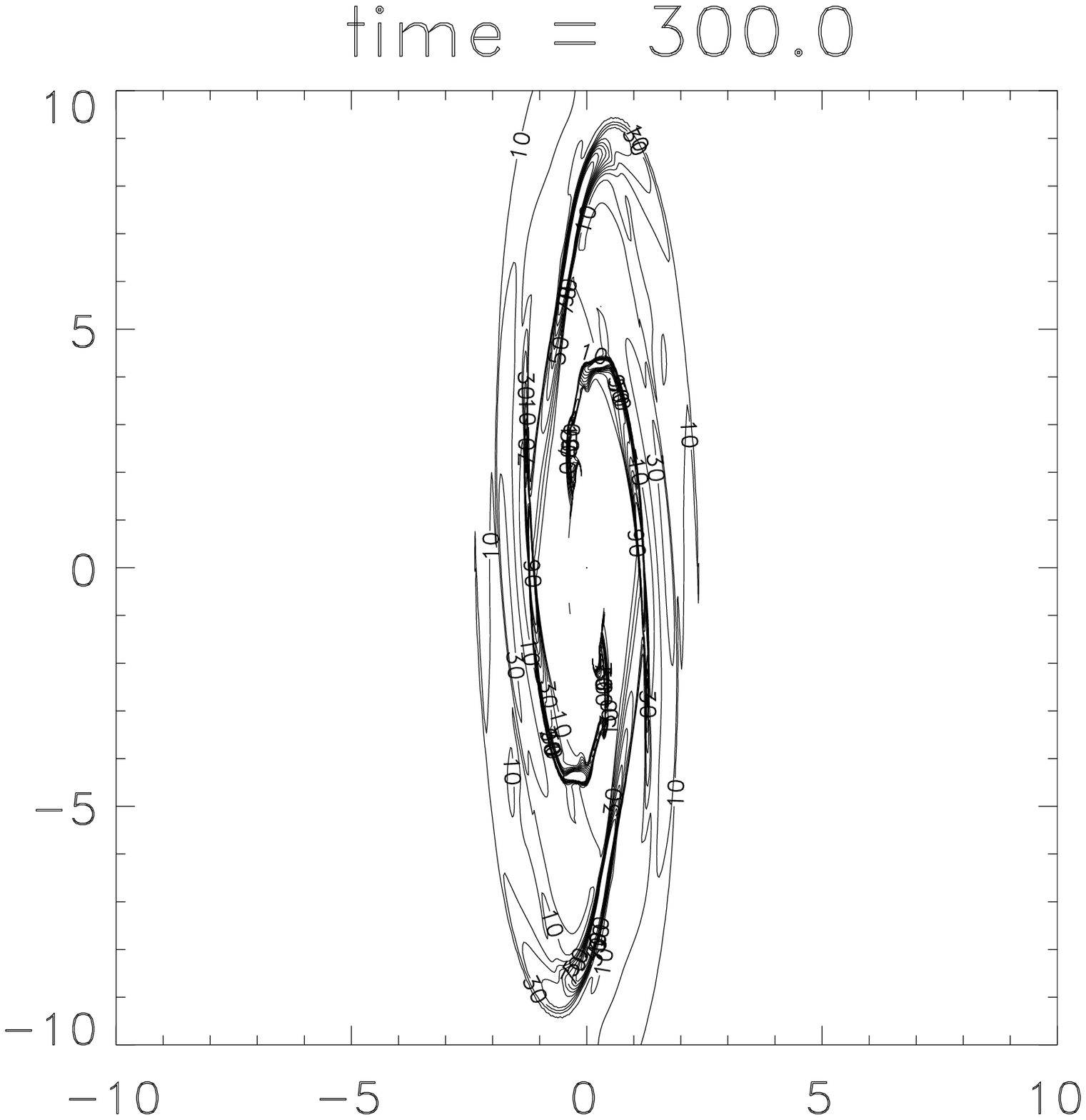}
     \end{minipage}
   \end{center}
   \caption{Left, B-band image of NGC~4650A with spiral structure indicated
      (Figure 10 of Arnaboldi et al.\ \cite{arnaboldi97});
       right, the surface density distribution 
       of Fig.\ \ref{m11_dens_t300} seen in projection, as if the
       disk is tilted about the y-axis.}       
   \label{n4650_m11_comp}
\end{figure}

 The temporal evolution of the global modes (Fig.\ \ref{m11_modes}) differs
significantly from the simulation without a central perturbation. 
The perturbed disk develops quickly linearily growing even modes ($m=2$ and 
$m=4$), whereas the odd modes remain on their initial low level. 
After a time $t \sim 120 \approx 1.8 \,\, {\rm Gyr}$ these 
modes become non-linear, which is reflected in an enhanced growth of the
$m=0$-mode. This radial mass redistribution is also shown by the 
Lagrange radii (Fig.\ \ref{m11_lagrange}).
The mass inside the 10\%-radius flows inwards, whereas the matter
up to the half-mass radius expands. At the time $t\sim 150 \approx 2.2$ Gyr
the $m=2$-mode reaches its (first) saturation level of about 10\%,
while the $m=4$-mode 
reaches a lower amplitude of $\sim 10^{-3} - 10^{-2}$. 
At about the same time the odd modes begin to grow, but they
are still linear at the end of the simulation at $t=300$. Therefore,
throughout the whole evolution the pattern looks mainly two-armed with
a slight modulation introduced by the $m=4$-mode (Figs.\ 
\ref{m11_denspert_t60} and \ref{m11_dens_t300}). 

It is remarkable, however, that the initial spiral pattern is leading
(as Fig.\ \ref{m11_denspert_t60} shows) whereas the final spiral
pattern is trailing like in the unperturbed model A. Moreover, the
occurence of an initial leading pattern is found in all our models
with a central S0 component. It is especially independent from boundary
conditions or other ''technical'' parameters.

\begin{figure}
   \resizebox{\hsize}{!}{
     \includegraphics[angle=90]{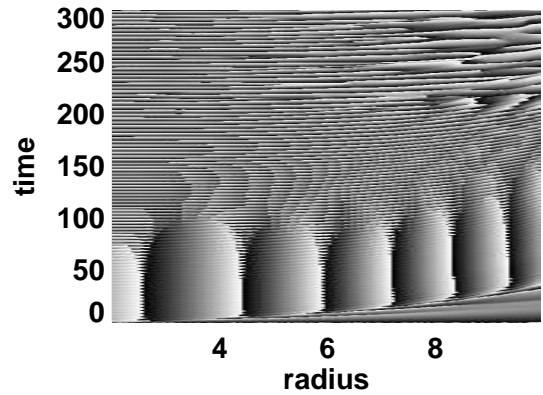}
   }
   \caption{Evolution of the phase angles for the $m=2$ mode
            (0$^\circ$: white; 360$^\circ$: black).}
   \label{m11_phaseangle}
\end{figure}

\begin{figure*}
   \begin{center}
     \begin{minipage}{15cm}
        \includegraphics[angle=0,width=7cm]{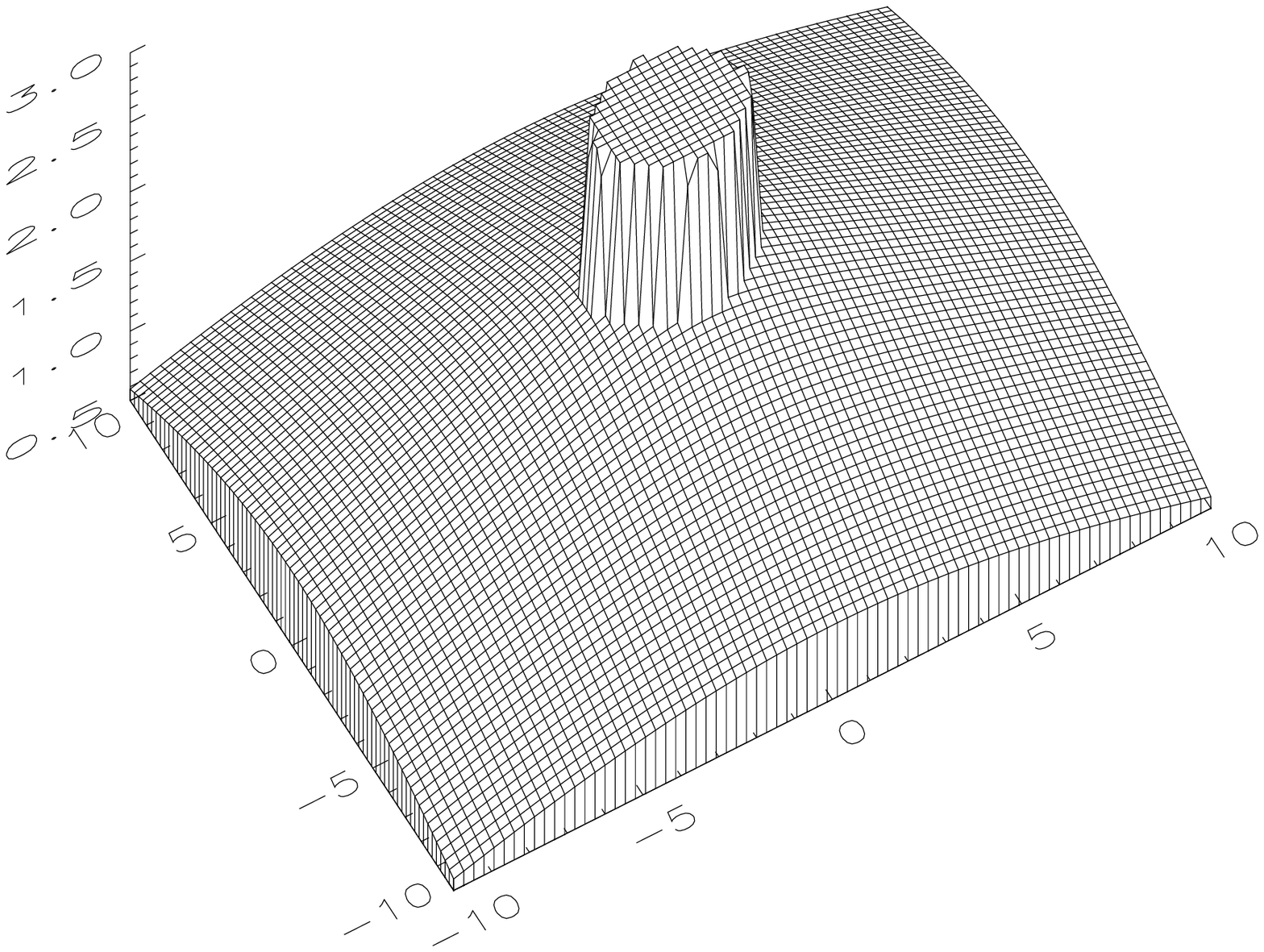}
        \includegraphics[angle=0,width=7cm]{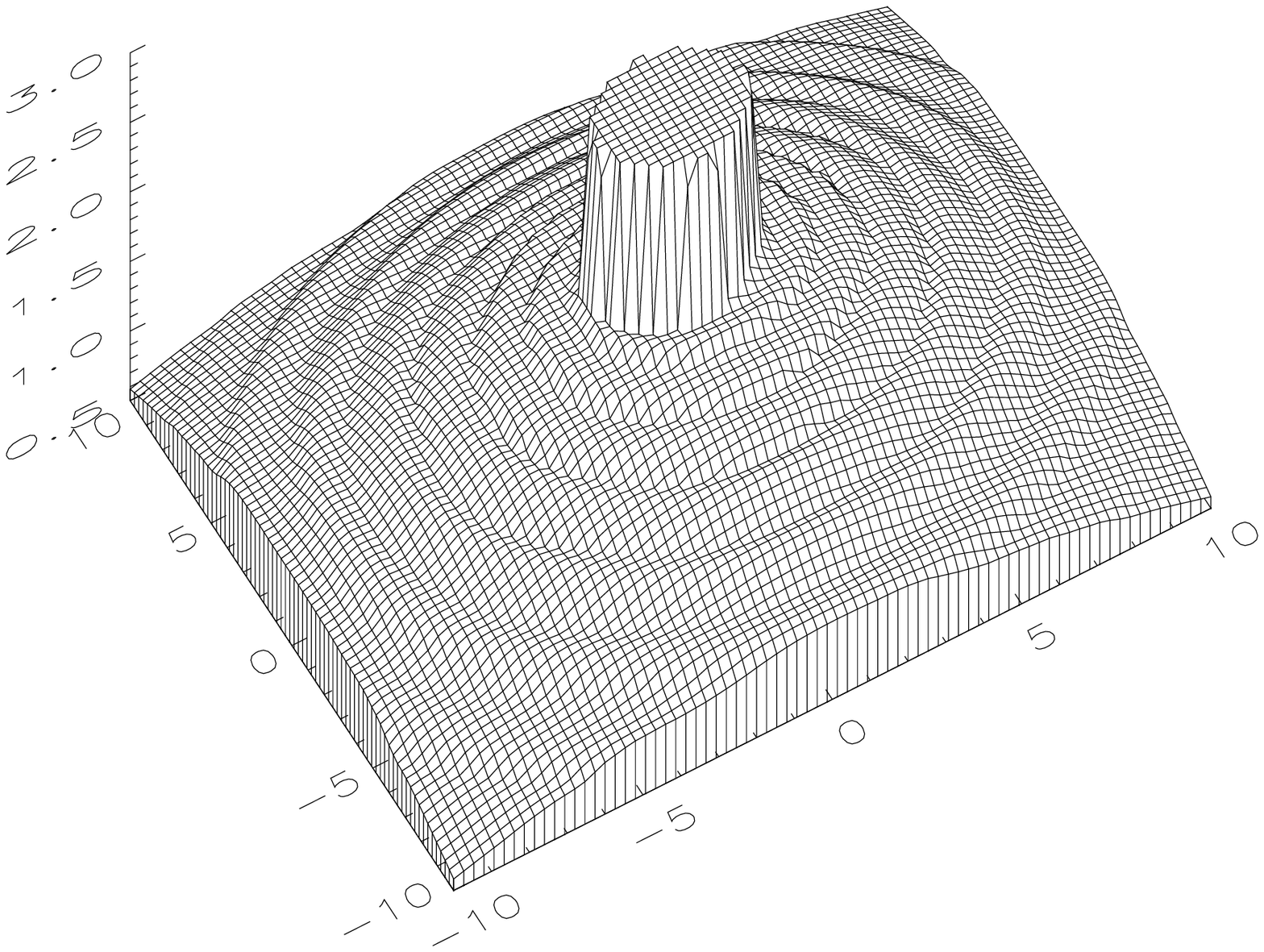}
     \end{minipage}
     \begin{minipage}{15cm}
        \includegraphics[angle=0,width=7cm]{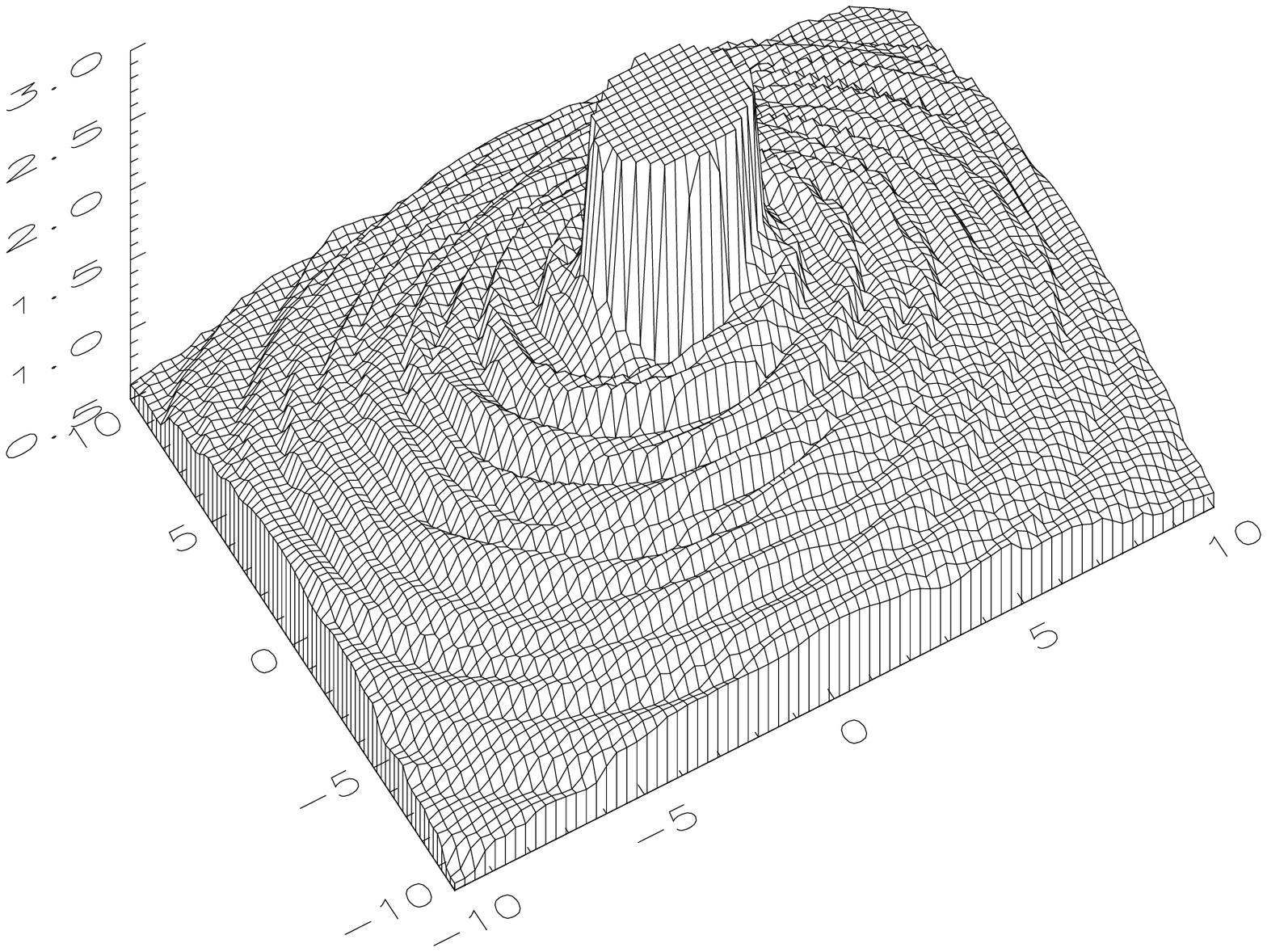}
        \includegraphics[angle=0,width=7cm]{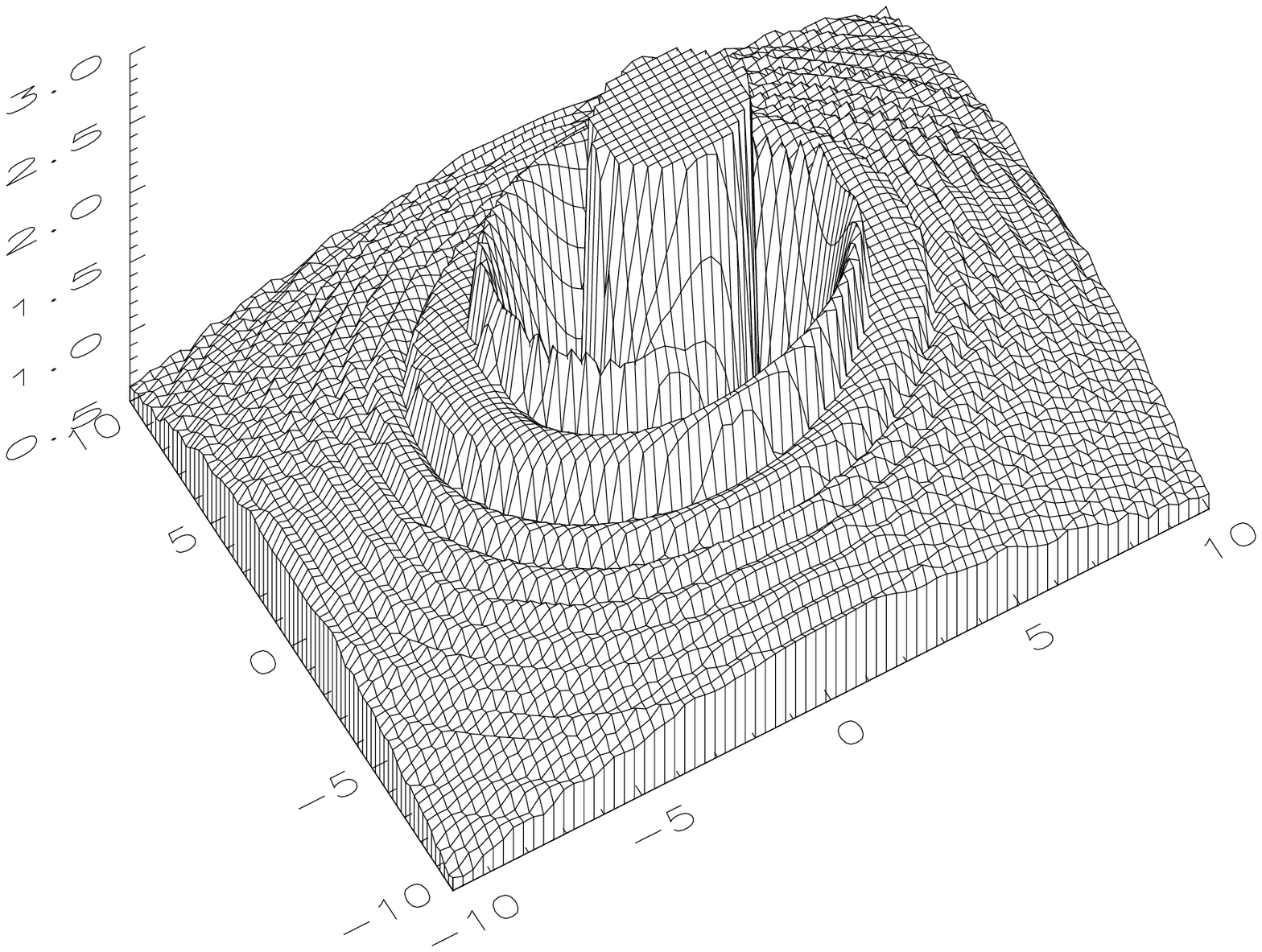}
     \end{minipage}
     \begin{minipage}{15cm}
        \includegraphics[angle=0,width=7cm]{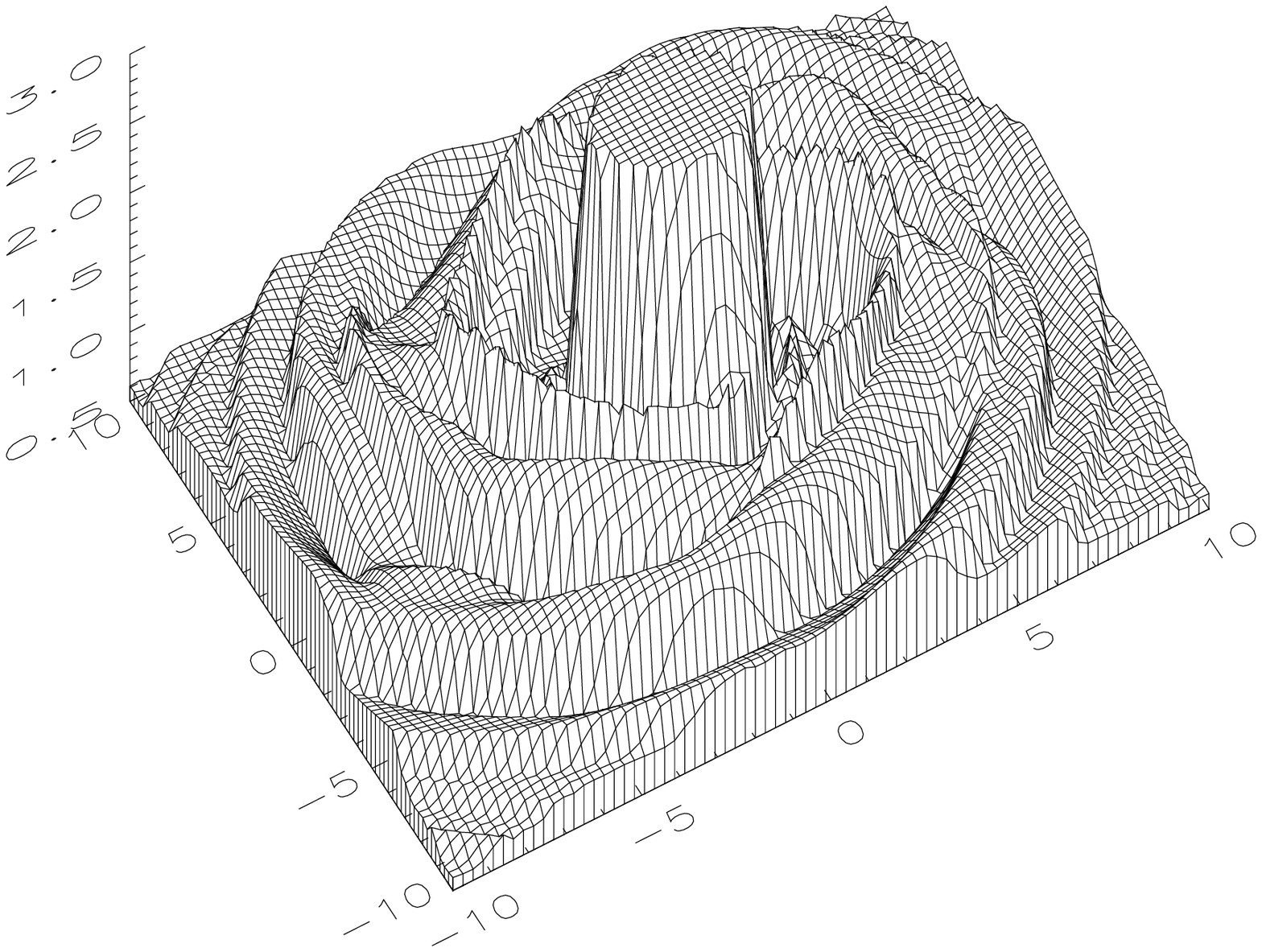}
        \includegraphics[angle=0,width=7cm]{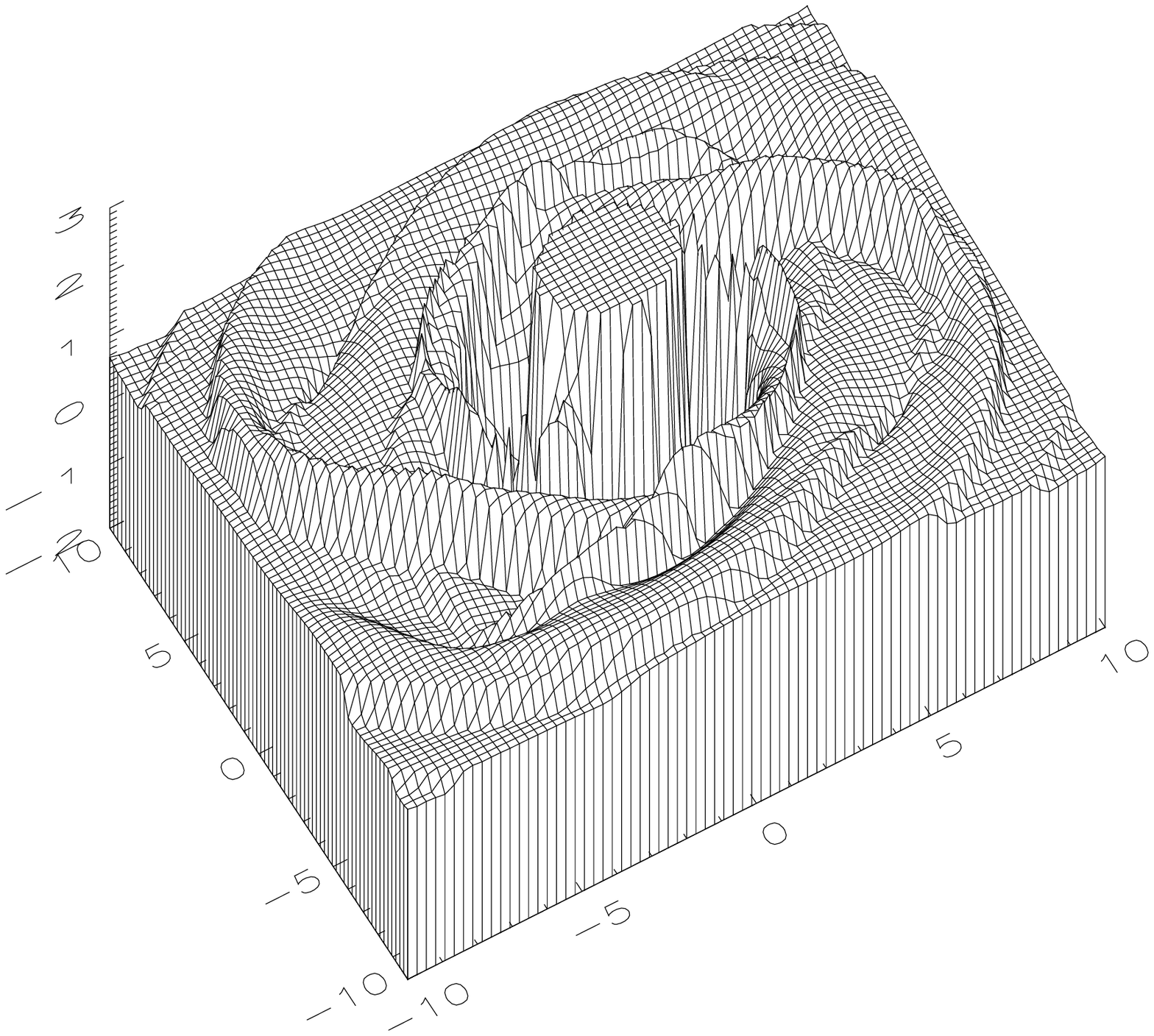}
     \end{minipage}
   \caption{Evolution of the logarithm of the surface density (in
            \mbox{$\msun$ pc$^{-2}$}) at different times 
            ($t=0$ ({\rm upper left}), $60$ ({\rm upper right}), $120$,
              $180$, $240$, $300$ ({\rm lower right})) for model B.
             For a better visualization of the ''punched'' central area 
             a cylinder with a constant level of 3.0 is drawn.}
   \label{m11_surfevol}
   \end{center}
\end{figure*}

The temporal evolution of the surface density is depicted in
Fig.~\ref{m11_surfevol}: during the linear stage (until $t=120$) a
regular tight leading spiral pattern emerges which becomes more and
more pronounced. Later, large density variations show up which make
the appearance temporarily ring-like. E.g.\ at $t=180$ there is an
almost closed ring at a radius of about 4 kpc. Towards the end of the
simulation a new quasi-equilibrium is established.  This new mass
distribution leads to more open trailing spiral arms in the saturation
phase (cf.\ also Fig.\ \ref{m11_dens_t300}).  When projected as if
seen close to edge-on, the spiral qualitatively resembles that seen in
NGC~4650A (Fig. \ref{n4650_m11_comp}).  We will show below that this
open spiral is essentially independent of the central S0 galaxy, which
acts only as a `seed' for its formation.  Obviously, the two-armed
perturbation created by the central bar-like S0 component
induces a mainly two-armed
structure in the disk. This becomes clear in the diagram of the phase
angles (Fig.\ \ref{m11_phaseangle}).  Immediately after the start of
the simulation a regular, but stationary pattern is formed from inward
out. This stationarity can be understood as the response of the flow
to the stationary central perturbation.  At a time $t \sim 90-100$
this large-scale tight leading pattern vanishes and a rapidly rotating
open trailing spiral is formed. The pattern speed (derived from the
temporal change of the phase angles) at that time is about \mbox{$0.9
  \approx 60$ km\,s$^{-1}$\,kpc$^{-1}$} which is close to, but lower
than the intrinsic pattern speed of the unperturbed disk.  The
transition from stationary pattern to rotating pattern occurs when the
azimuthal force of the growing spiral structure begins to exceed the
azimuthal force of the central component. They become comparable for
the first time at about $t \sim 85$. Afterwards the azimuthal force of
the spirals quickly outnumbers the contribution of the 
bar-like central component, making the central perturbation 
unimportant for the further growth of the spiral structure.


\subsection{Variation of the central component}

  In order to test the influence of the central component we varied the
mass of the central component (models MB1 and MB2), its switch-on 
characteristics (model T1) as well as the duration of the central
perturbation (model T2). 

\begin{figure}
   \resizebox{\hsize}{!}{
     \includegraphics[angle=270]{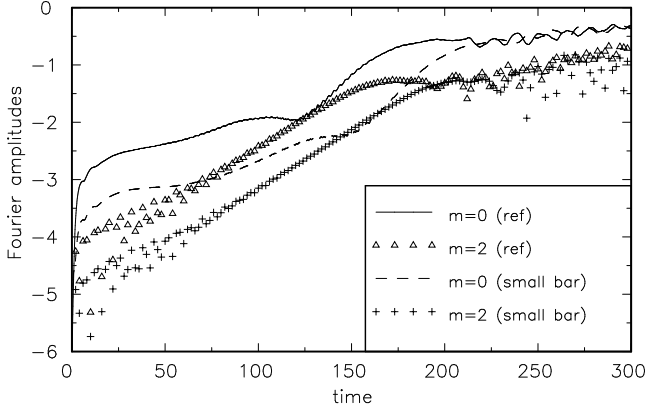}
   }
   \caption{Temporal evolution of the amplitudes of the $m=0,2$-modes
      for the reference model B and a small central component 
      ($10^9 \msun$, model MB1).}
   \label{m11_smallbar}
\end{figure}

\begin{figure}
   \resizebox{\hsize}{!}{
     \includegraphics[angle=270]{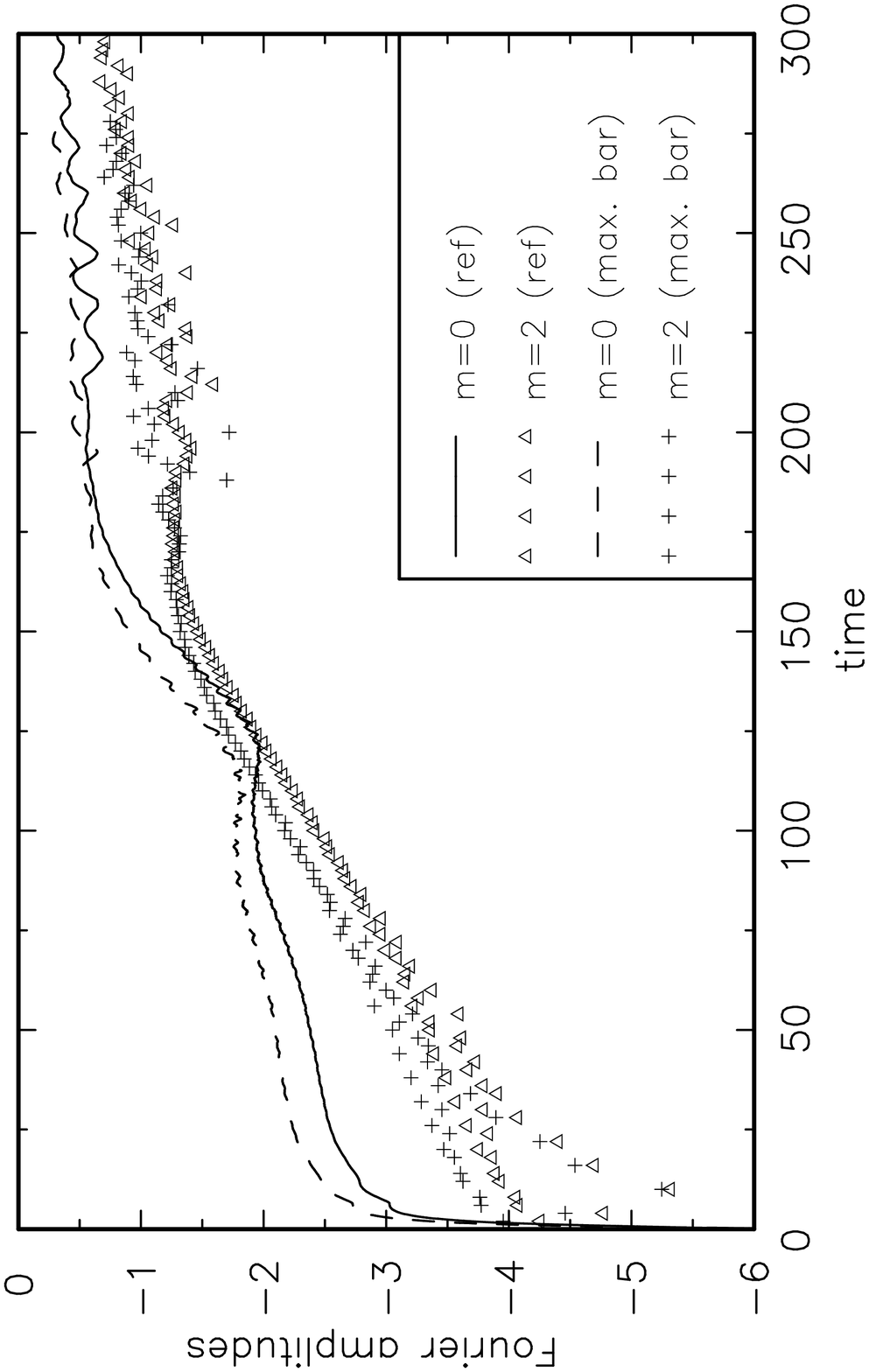}
   }
   \caption{Temporal evolution of the amplitudes of the $m=0,2$-modes
      for the reference model B and a maximum central
      component ($10^{10} \msun$, model MB2).}
   \label{m11_maxbar}
\end{figure}

\begin{figure}
   \resizebox{\hsize}{!}{
     \includegraphics[angle=270]{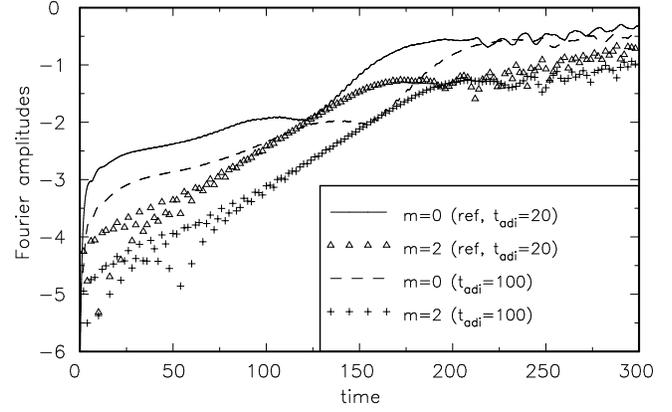}
   }
   \caption{Temporal evolution of the amplitudes of the $m=0,2$-modes
      for the reference model B and a central component growing slower
      by a factor of 5 (model T1).}
   \label{m11_aditime}
\end{figure}

{{\bf Mass of central component.}  
    Simulations with a bar-like S0 component
  that is five times less massive ($1 \cdot
  10^9 \,\, \msun$: model MB1) and with a 
  ''maximum S0 component'' of $10^{10}
  \,\, \msun$ (model MB2) have been performed.  The smaller mass leads
  to a delayed growth of the spiral modes of the disk by about 
  \mbox{$t \sim 25 \approx 370$ Myr} (Fig.\ \ref{m11_smallbar}). In
  case of the more massive central pertubation there is 
  again a small temporal shift
  in the behaviour of the amplitudes during the linear stage (Fig.\ 
  \ref{m11_maxbar}). The amplitudes run now ahead by about 100-150
  Myr. In both cases the growth rates and the final saturation levels
  are unaffected by the varied mass of the central component.

  We also varied the density profile of the central component by
choosing $n_b=-1$ (MB3) and $n_b=-2$ (MB4). However, even the steeper 
profile did not change the numerical results significantly.

{\bf Timescale of perturbation due to central component.}
Another test of the influence of the central component is provided
by a different temporal behaviour of the perturbing bar-like potential.
In model T1 the switch-on time $t_{\rm adi}$ was increased by a factor
of 5 to $t_{\rm adi}=100$. Similar to the model MB1 with a less massive 
central perturbation the growth of the main amplitudes is now
delayed by about 400 Myr (Fig.\ \ref{m11_aditime}). The growth rates 
and the saturation levels, however, remain unaffected.


\begin{figure}
   \resizebox{\hsize}{!}{
     \includegraphics[angle=270]{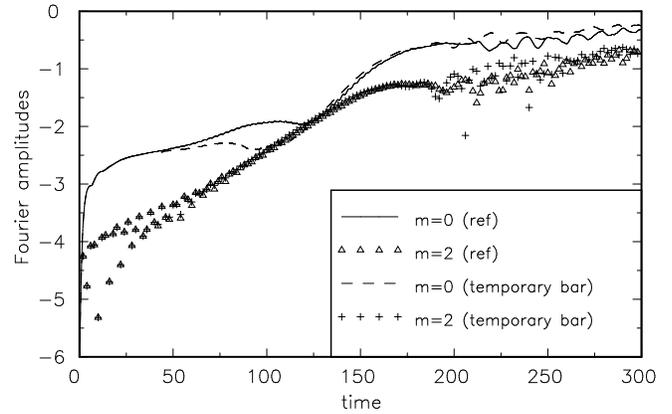}
   }
   \caption{Temporal evolution of the amplitudes of the $m=0,2$-modes
      for the reference model B and model T2 in which the central component 
      is switched off between $t=80$ and $t=100$ (1.2-1.5 Gyr).}
   \label{m11_temporarybar}
\end{figure}

Just as in the previous models, even a complete removal of the 
S0 component
does not alter the evolution strongly.  Fig.\ref{m11_temporarybar}
shows that the disk quickly evolves on its standard evolutionary path
ending up in the same saturation stage, even if the S0 component is switched
off before the density perturbations in the disk reach the non-linear
regime.  The perturbation induced by the S0 component does not
disappear when the asphericity of the central perturbation is switched
off, and the modes continue to grow.  When the gaseous component is
dynamically substantially hotter, we find that the excited modes are
erased after the perturbation is switched off.

  From these calculations we conclude that the modal growth in the
reference model B decouples from 
the initial central perturbation after time $t \sim 100$. The central 
component acts as a seed for the growing modes, but once 
these modes are excited, they decouple from the seed growing
on a timescale of 1-2 Gyrs. Also other perturbations, e.g.\ due to gas
accretion, might act as such a seed.


\subsection{Mass of the polar disk}


\begin{figure}
   \resizebox{\hsize}{!}{
     \includegraphics[angle=270]{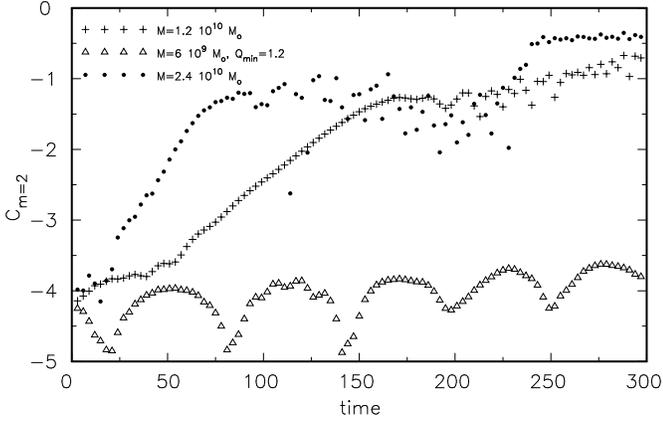}
   }
   \caption{Temporal evolution of the amplitudes of the $m=2$-modes
      for the reference model B ($1.2 \cdot 10^{10} \msun$, plus) and 
      models with different disk masses:
      MD2 ($6.0 \cdot 10^{9} \msun$, triangle) and 
      MD3 ($2.4 \cdot 10^{10} \msun$, filled circle).}
   \label{amp_diskmass}
\end{figure}

    The influence of the mass of the polar disk is studied
by a small series of simulations starting from half of the mass 
of the reference model (MD1, MD2) up to the maximum polar disk mass of
$\natd{3.7}{10} \msun$ (MD4). A disk with half the mass of 
the reference polar disk (or about 1/6 of the maximum polar disk) 
is very stable (MD1): after an initial adjustment the
amplitudes of all modes remain almost constant on a very low level. 
The dominant $m=2$ and $m=4$ modes are still deep in the linear
regime. Only a weak and very tightly wound leading spiral is formed similar to 
that found in the early stage of the reference model. Model MD2
has the same lowered disk mass as model MD1, but a reduced minimum Toomre 
parameter $Q_{\rm min}=1.2$. Still the disk is rather stable 
(Fig.\ \ref{amp_diskmass}) and the
dominant even modes have the same amplitude as in MD1. The main
difference to MD1 is that the odd modes rise very slowly, but after 
4.5 Gyr their amplitudes are still well below those of the even modes.

  Increasing the disk mass to twice the mass of the reference model,
destabilizes the polar disk (MD3, Fig.\ \ref{amp_diskmass}). 
The growth rates of the $m=2$
and $m=4$ modes are doubled compared to the reference model. The growth
rates of the $m=1$ and $m=3$ modes are very similar to those
of the even modes. The overall appearance within the first
2 Gyr, however, is dominated by the even modes because their early 
amplitude is raised by the perturbation due to the central 
S0 component. Increasing the mass up to the maximum polar disk (MD4)
gives a result very similar to the model with doubled mass MD3.


\subsection{The Toomre parameter and the rotation curve}

\begin{figure}
   \resizebox{\hsize}{!}{
     \includegraphics[angle=270]{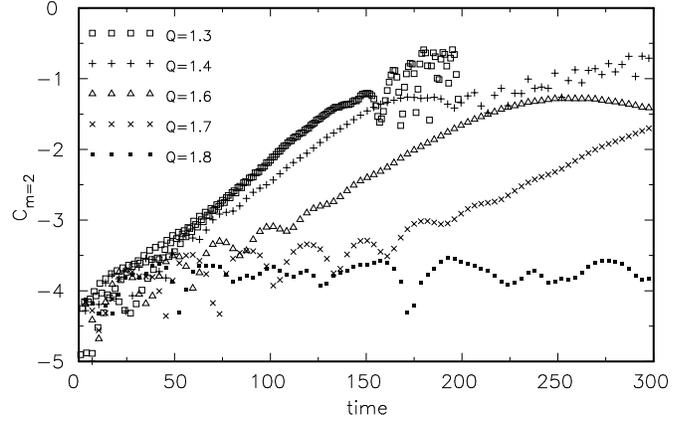}
   }
   \caption{Temporal evolution of the amplitudes of the $m=2$-modes
      for the reference model B and models Q1 \dots Q6 with different 
      initial minimum Toomre parameters (Q=1.3,1.6,\dots,1.9).}
   \label{m11_Qdependence}
\end{figure}

An important parameter for the stability of disks is the 
Toomre parameter. In a series of models we varied the initial
minimum $Q$-value between 1.3 and 1.9. Our reference model
B was characterized by $Q_{\rm min}=1.4$. Fig.\ \ref{m11_Qdependence}
demonstrates the strong dependence of the growth rate on $Q$ during the linear
stage: hotter disks are more stable. Close to the reference model
(e.g.\ comparing models with $Q=1.3$ and $Q=1.5$) the growth rate varies 
less strongly than in models with larger $Q$-values (e.g.\ the range 
$Q=1.5$ to $Q=1.7$). 
Moreover, the disks show a transition to complete stability when the
Toomre parameter exceeds a value between 1.7 and 1.8. 
This is in good agreement with analytical predictions 
for the transition to stability to be at $Q_c \approx 1.73$ 
(in case of flat rotation curves, see e.g.\ GB00). 
The saturation levels of the dominant $m=2$-modes are -- in case of unstable 
configurations -- rather similar, i.e.\ of the order of about 10\%. 

In the models R1 and R2 we tested the influence of a broader transition
region between the rigid rotation domain and the flat part of 
the rotation curve by setting $n=6$ (R1) and $n=3$ (R2). 
Model R1 develops qualitatively similar to the reference model B.
However, the growth rates of the modes are 30\% lower and,
therefore, the non-linear regime is first reached after about 3.5 Gyr.
In model R2 the transition region is even broader. Still a
growing instability is found, but it grows on a longer timescale. The
structures dominated by the lower even modes become non-linear at about 
$t \approx 300 \approx 4.5 \, \mathrm{Gyr}$. 
Following the linear stability analysis of 
Polyachenko et al.\ (\cite{polyachenko97})
the critical Toomre parameter $Q_c$ (for which the system is stable
against all perturbations; see their Eq.\ (36)) 
drops from $Q_c = \sqrt{3} \approx 1.73$ for flat rotation curves
to $Q_c = \sqrt{3/2} \approx 1.22$ for the rigid rotation region.
Thus, a broader transition region results in a reduced $Q_c$ beyond the
transition radius $R_\mathrm{flat}$ compared to the reference model,
and the system is expected to be more stable, in agreement 
with our simulations.


\subsection{Equation of state}

\begin{figure}
   \resizebox{\hsize}{!}{
     \includegraphics[angle=270]{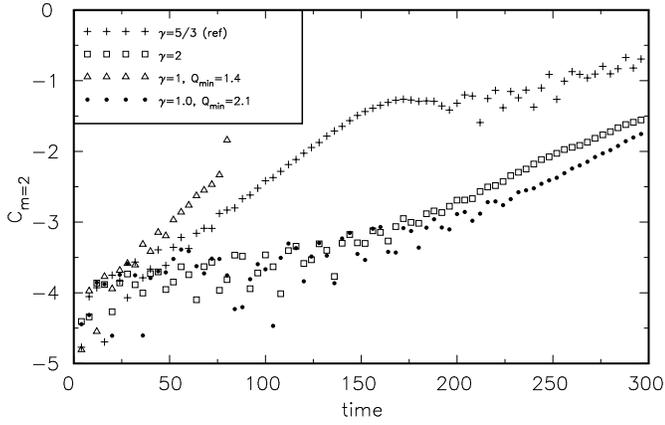}
   }
   \caption{Temporal evolution of the $m$=2-amplitude
     for different equations of state: $\gamma$=5/3 (reference model
     B, plus sign), $\gamma$=2 (model S1, open box) and $\gamma$=1
     with $Q_\mathrm{min}$=1.4 (model S2, open triangle) and
     $Q_\mathrm{min}$=2.1 (model S3, filled circle).}
   \label{m11_eqofstate}
\end{figure}

In the models S1 to S3 we changed the equation of state by varying the
adiabatic exponent $\gamma$ in the polytropic equation of state.  In
model S1 $\gamma$ was set to 2 which is often adopted to mimic stellar
disks (e.g.\ Orlova et al.\ \cite{orlova02}). Still such a disk is not
stable, however the growth of the Fourier amplitudes is delayed and
smaller (Fig.\ \ref{m11_eqofstate}).  Such a stabilizing behaviour can
be understood from the stiffer equation of state for $\gamma=2$ which
hinders perturbations from growing. 

Correspondingly, the disk is expected to be more unstable when the
equation of state is less stiff as e.g.\ in the isothermal case
$\gamma=1$.  
Here cooling is taken to be so efficient that all the heat
  of compression is dissipated immediately, so that the disk is
  maintained in an isothermal state, as suggested by Lodato \& Rice (2005).
Accordingly, we see that model S2
reaches the non-linear regime in about half the time of
the reference model. However, the destabilizing effect of the reduced
stiffness can be partly compensated for by an initially hotter disk, as
model S3 with a minimum Toomre parameter of $Q_\mathrm{min}=2.1$
demonstrates.  
The cases $\gamma = 1, 5/3$ and 2 are expected to bracket the 
expected cooling behaviour in real disks.


\subsection{Miscellaneous}

\begin{figure}
   \resizebox{\hsize}{!}{
     \includegraphics[angle=270]{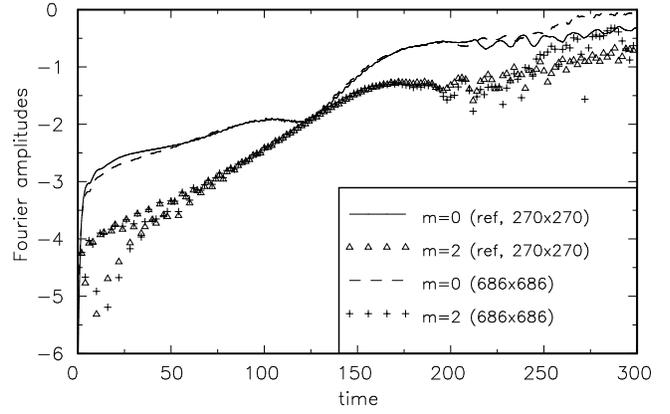}
   }
   \caption{Temporal evolution of the amplitudes of the $m=0,2$-modes
      for the reference model B (270x270) and a model G1
      with increased grid size (686x686).}
   \label{m11_gridsize}
\end{figure}

    We tested different variations of ''technical'' parameters which
did not result in substantial quantitative deviations from the
reference models. E.g.\ we varied the numerical timestep criterion, we
compared with models using different amounts of artificial viscosity
or we changed the grid size. As an example Fig.~\ref{m11_gridsize}
demonstrates the insensitivity of the numerical results on the grid
size. Though the resolution was more than doubled, the amplitudes
characterizing the dominant mode or the radial mass redistribution
evolve qualitatively in the same way and quantitatively almost
identical.

   In order to test the influence of the inner boundary we rerun
the unperturbed model A, but with an inner edge at 
$R_\mathrm{min} = 1 \mathrm{kpc}$ (models G2 and G3). In both
models we kept the total disk mass constant. In model G2 the exponential
disk was extended down to a radius of 1 kpc, while in model
G3 the surface density was exactly exponential outside
a radius of 2 kpc (like in model A), but had a smooth transition
to zero from a radius of 2 kpc down to the inner edge. 
Both models were more stable than model A. In model G2 the
growth rate of the dominant $m=1$-mode is a factor two smaller
than the growth rate of the dominant mode $m=2$ in model A.
Model G3 shows no significant growth at all. In model G4 the
same smooth surface density distribution as in model G3 was used, but the
minimum Toomre parameter was reduced to 1.2. In that case, the
model becomes violently unstable after 4 Gyr. This means that the
stabilizing effect of a smoother mass distribution can be
compensated by a slightly reduced minimum Toomre parameter.
Taking into account that the velocity dispersions within the polar disk
are poorly known (if at all), the uncertainties related to the mass 
distribution at the inner edge of the disk are less critical, 
unless the profile of the Toomre parameter can be constrained 
in more detail.


\section{Discussion and summary}
\label{discussion}

We investigated the stability properties of polar disks by
2-dimensional hydrodynamical simulations of flat disks perturbed by a
non-rotating bar-like potential mimicking a central S0-like component. 
The properties of
the polar disk and the central perturbation were derived from
observations of the prototypical polar ring galaxy NGC4650A. We
modelled the polar ring by a disk with a ``punched'' central hole and
an exponential mass distribution embedded in a halo with a mainly flat
rotation curve.  We modelled the polar gas using a polytropic
equation of state (EOS), and did not consider a multi-phase ISM or
energy feedback from star
formation in our calculations. Both will probably influence especially
the late non-linear stages of the evolution, when cool clumps can form
and star formation sets in.  E.g.\ star formation is expected to
stabilize the disk due to its energy feedback to the ISM, but a
population of cool dense clouds is destabilizing (e.g.\ Jog \& Solomon
\cite{jog84}, Shen \& Lou \cite{shen04}).  However, the timescale for
the disk to become unstable should not be affected strongly.
In order to study different kind of gas behaviour we varied the
adiabatic exponent covering the isothermal case ($\gamma=1$) up to
the very stiff ''stellar'' case ($\gamma=2$). Though the numerical
results depend on the actual EOS, its influence also strongly depends
on the Toomre parameter of the disk, in the sense that a hotter disk 
can compensate for a less stiff EOS.

Without the central perturbation, the disk is fairly stable over the
whole simulation time of 4.5 Gyr. Two-armed and four-armed modes grow,
but their growth rates are rather small. The corresponding Fourier
amplitudes of $10^{-4}$ are even after 4.5 Gyr deeply in the linear
regime and, therefore, basically undetectable to observation.

Exciting the perturbation by a central bar-like component representing
the S0 galaxy changes the evolution of the polar disk drastically.
Though the perturbation was switched on adiabatically, the Fourier
amplitudes reach a level of $10^{-4}$ more or less as soon as the 
S0 potential is switched on.  Tightly wound leading spirals are formed;
though their flow pattern remains stationary, their amplitude is
growing.  In our simulation the azimuthal component of the
gravitational force exerted by the spiral perturbations becomes
comparable to that of the central S0 at about 1 Gyr.  The evolution of
the polar disk then begins to decouple from the central component.  At
about 1.8 Gyr a substantial radial mass redistribution sets in,
indicating that the evolution becomes non-linear. The tightly wound
spiral disappears and, finally, a much more open, but rotating,
two-armed trailing spiral develops.

The general behaviour turned out to be rather robust against
variations of the central component (e.g.\ its mass or the temporal
behaviour of the perturbation).  Switching off the non-axisymmetric
component of the imposed external potential, even if it is done well
before the nonlinearity has developed, has no apparent effect on the
state at the end of the simulation.

The most important parameter for the evolution of the polar disk is
the Toomre parameter. In case of a high $Q$ exceeding 1.7, only
tightly wound spirals form,
driven by the barlike perturbation of the S0 disk.
Such disks are too hot to be unstable
against the induced spiral perturbations, and the tight spirals
disappear when the central component is switched off. 
This is exactly the reason usually given for the absence of spiral
structure in the stellar disks of S0 galaxies: the disk is too hot to
respond coherently and support a self-excited global spiral mode
(e.g.\ Chapter 18 of GB00).
In case of lower $Q$ (1.6 or less) the central component just acts as
a seed of instability. In the early stage a tightly wound leading
spiral is formed like in the reference model. This has a strong
resemblance to the spiral in the inner polar disk of NGC~2787 (see
e.g.\ Erwin \& Sparke \cite{erwin03}).

Later on, the spiral structure decouples and becomes an open trailing
spiral during the saturation stage. The timescale for reaching the
saturation stage slightly depends on $Q$ ranging from 2.2 Gyr
($Q=1.3$) to 3 Gyr ($Q=1.6$).  Taking a Fourier amplitude for the
$m=2$-mode of 1\% as the margin for a detectable spiral structure, the
timescale for our reference model of a cold polar disk to become
unstable is about 1-2 Gyr. The transition to the saturation stage
takes then another Gyr. These timescales are remarkably independent of
the mass of the central component, and depends mainly on the mass in
the polar disk and the Toomre parameter.  If the mass is halved, the
instability disappears unless the Toomre parameter is lowered, and the
trailing spiral saturates only after 4.5~Gyr.  If the polar-disk mass
is doubled, growth rates are also approximately doubled, but the
saturated spiral looks similar.  
Again, studies of stellar systems have long noted that 
relatively cool disks containing a large fraction of the system's mass
are likely to be unstable to forming a spiral pattern: see e.g\ 
Binney \& Tremaine (\cite{binney87}) for a summary.
The spirals formed in that stage are
all trailing. The open spiral in NGC~4650A described by Arnaboldi et
al.\ (\cite{arnaboldi97}) and Gallagher et al.\ (\cite{gallagher02})
might be an example of a late stage of such an evolution.

On the basis of these computations, we expect an initially
axisymmetric polar disk resembling that in NGC4650A to be stable at
least over 1-2 Gyr, or longer if it is hot.  When they become
unstable, they first form tightly wound leading spirals, before they
rearrange their mass building up an open trailing spiral.  Hence, the
variety of observed structures in polar disks might imply a large
range of ages.

In order to predict the long-term dynamical evolution of a polar disk,
it is essential to determine its Toomre parameter. Hence, in addition
to the rotation curve and the mass distribution, the velocity
dispersions must be known.

Additionally, it would be very interesting to test whether observed
polar disks generally have some substructure and if so, how many of
them show tightly wound or open spirals. Moreover, it would be very
important to check whether there are tightly wound spirals which are
indeed leading.  Such a measurement, however, is very difficult due to
the small Fourier amplitudes in the stage when leading arms are
expected.  In the transition stage when the leading spirals become
trailing, the arms are still tightly wound. Hence tightly bound arms
can be trailing or leading depending on their evolutionary stage.

Another major problem is that polar disks are most easily visible when
they are almost edge-on. However, then it is difficult to analyse the
substructure in the disk (see e.g.\ Figs.\ 3a-3d in Whitmore et al.'s
(\cite{whitmore90}) catalog of polar-ring galaxies). However, deep
images of a selected sample of inclined polar disks might allow for
such a measurement.

Bekki \& Freeman (\cite{bekki02}) have given an alternative
explanation for the spiral in the polar disk of NGC~4650A.  The
perturbation by a triaxial dark matter halo tumbling slowly about an
axis perpendicular to the polar ring can cause an open spiral
structure well outside the region where we expect the polar disk to be
self-gravitating.  They emphasize that self-gravity is less important
for their structure formation mechanism, because they produce mainly
two-armed kinematic spirals.  However, the rotation of the dark halo
must be finely tuned to produce the observed open spiral.

We have demonstrated that the (unavoidable) perturbation given by the
central (S0) component of a polar ring galaxy embedded in a spherical
dark matter halo component can be sufficient to create open trailing
spiral arms well outside the central S0 galaxy.  The self-gravity of
our polar disk is essential for creating pronounced spiral structures
with amplitudes in the non-linear regime, which need not necessarily
to be dominated by even modes.  On the other hand, in low mass polar
disks it is very difficult to form outer spirals by a central
perturbation.  We may see a tight trailing spiral forced by the
perturbation of the central oblate galaxy, or no spiral pattern at
all.

\begin{acknowledgements}
  Part of this work was supported by the German Science Foundation
  \emph{DFG}, project number TH511/6 (CPT), by the US National Science
  Foundation under grants AST-00-98419 (LSS), and AST-9803018 (JSG).
  We are grateful to the Computing Centre of the University of Kiel,
  where most of the simulations have been performed. CPT is grateful
  for the hospitality of the Dept.\ of Astronomy of the Univ.\ of
  Wisconsin-Madison.
\end{acknowledgements}


\end{document}